\documentclass[11pt]{amsart}%
\usepackage{amsmath}
\usepackage{amsfonts}
\usepackage{amssymb}
\usepackage{graphicx}%
\providecommand{\U}[1]{\protect\rule{.1in}{.1in}}
\newtheorem{theorem}{Theorem}

\newtheorem{lemma}[theorem]{Lemma}

\email{chams@aub.edu.lb}
\email{alain@connes.org}
\begin{document}
\begin{titlepage}
\vspace{.3cm} \vspace{1cm}

\begin{center}
\baselineskip=16pt \centerline{\Large\bf Noncommutative Geometric Spaces}
\vspace{.5cm} \centerline{\Large\bf with Boundary: Spectral Action }
\vspace{2truecm} \centerline{\large\bf Ali H.
Chamseddine$^{1,3}$\ , \ Alain Connes$^{2,3,4}$\ \ } \vspace{.5truecm}
\emph{\centerline{$^{1}$Physics Department, American University of Beirut, Lebanon}}
\emph{\centerline{$^{2}$College de France, 3 rue Ulm, F75005, Paris, France}}
\emph{\centerline{$^{3}$I.H.E.S. F-91440 Bures-sur-Yvette, France}}
\emph{\centerline{$^{4}$Department of Mathematics, Vanderbilt University, Nashville, TN 37240 USA}}%

\end{center}

\vspace{2cm}

\begin{center}
\textbf{Abstract}
\end{center}

We study spectral action for Riemannian manifolds with boundary, and then
generalize this to noncommutative spaces which are products of a Riemannian
manifold times a finite space. We determine the boundary conditions consistent
with the hermiticity of the Dirac operator. We then define spectral triples of
noncommutative spaces with boundary. In particular we evaluate the spectral
action corresponding to the noncommutative space of the standard model and
show that the Einstein-Hilbert action gets modified by the addition of the
extrinsic curvature terms with the right sign and coefficient necessary for
consistency of the Hamiltonian. We also include effects due to the addition of
dilaton field.
\end{titlepage}\bigskip
\tableofcontents

\section{Introduction}

Boundary of manifolds play an important role in many physical theories, such
as anomalies, Chern-Simons theories, topological theories, conformal theories
and gravity. Riemannian geometry of manifolds with boundary are well
understood. This is not the case in noncommutative geometry where the spectral
triple associated with the boundary of noncommutative space has not been
defined. The spectral action principle in noncommutative geometry states that
the physical action depends only on the spectrum. In particular, the simple
assumption that space-time is a product of a continuous manifold times a
finite space of $KO$-dimension $6$ results uniquely in the noncommutative
space of the standard \ model, predicting the number of fermions to be $16$
and determines the correct representations of these fermions with respect to
the gauge symmetry group $SU(3)\times SU(2)\times U(1)$ \cite{Why},
\cite{Beggar}. The spectral action is defined as the trace of an arbitrary
function of the Dirac operator for the bosonic part and a Dirac type action
for the fermionic part including all their interactions. The action is then
uniquely defined and the only arbitrariness one encounters is in the first few
moments of the function which enter in the spectral expansion with the higher
coefficients suppressed by the high-energy scale. One essential point in the
analysis is that the formulation is defined in terms of operators of compact
resolvent, and thus the space considered is Euclidean. Therefore the model
thus defined will correspond to Euclidean quantum gravity and will need Wick
rotation to go to spaces with Lorentzian signature. Space-time is then assumed
to have the topology of $\Sigma\times R$ where $\Sigma$ is a three dimensional
space. In studying the dynamics of the gravitational field by performing the
$3+1$ splitting, one discovers that the Hamiltonian obtained from the
Einstein-Hilbert action, contains additional unwanted surface term that could
be eliminated exactly by adding a surface term equal to twice the extrinsic
curvature of the boundary three space. Alternatively, the variation of the
Einstein-Hilbert action is inconsistent for manifolds with boundary without
the addition of the extrinsic curvature term. The question we have to face is
whether the spectral action for manifolds with boundary gives the correct
boundary terms. This will be a severe test on the spectral action principle,
because the boundary terms are completely fixed and there is essentially no
freedom allowed to change any of these terms. The plan of this paper is as
follows. In section 2 we summarize properties of Riemannian manifolds with
boundary. In section 3 we evaluate the spectral action for a Dirac operator on
a Riemannian manifold with boundary. In section 4 we define the spectral
triple associated with a noncommutative space with boundary. In section 5 we
evaluate the spectral action for the noncommutative space of the standard
model taken to be with boundary. In section 6 we include the effects of a
dilaton. Section 7 is the conclusion. Appendix A is our calibrating example
where the manifold is taken to be the disk, and this can be used to check the
sign conventions. Appendix B is a summary of the variation of the
Einstein-Hilbert action in presence of the boundary term. The results in this
paper were announced in \cite{quantumactions}.

\section{Riemannian manifolds with boundary}

We shall first give some definitions concerning embedding of hypersurfaces in
a manifold that will enable us to perform the computations in a covariant way.
Let us denote the coordinates of the manifold $M$ by $\left\{  x^{\mu
}\right\}  $ and of the hypersurface by $\left\{  y^{a}\right\}  $ and define
the unit \emph{inward} normal to the hypersurface by $n^{\mu}$ such that
$g_{\mu\nu}n^{\mu}n^{\nu}=1$ where $g_{\mu\nu}$ is the metric on $M$ which is
assumed to be Euclidean. Define the functions $e^{\mu}\left(  y^{a}\right)  $
as the embedding of the hypersurface in $M$ and let
\begin{equation}
e_{a}^{\mu}=\frac{\partial e^{\mu}}{\partial y^{a}}%
\end{equation}
then the metric $g_{\mu\nu}$ on $M$ induces a metric $h_{ab}$ on the
hypersurface such that
\begin{equation}
h_{ab}=g_{\mu\nu}e_{a}^{\mu}e_{b}^{\nu}%
\end{equation}
and where the inward normal $n^{\mu}$ is orthogonal to $e_{a}^{\mu}$
\begin{equation}
g_{\mu\nu}n^{\mu}e_{a}^{\nu}=0.
\end{equation}
It is convenient to define $n_{\mu}=g_{\mu\nu}n^{\nu}$ so that $n_{\mu}%
e_{a}^{\mu}=0.$ We now define the inverse functions $e_{\mu}^{a}$ by
\begin{equation}
e_{a}^{\mu}e_{\mu}^{b}=\delta_{a}^{b}%
\end{equation}
which satisfies the two conditions
\begin{equation}
e_{a}^{\mu}e_{\nu}^{a}=\delta_{\nu}^{\mu}-n^{\mu}n_{\nu}\,,\ \ n_{\mu}%
e_{a}^{\mu}=0.
\end{equation}
We therefore can write
\begin{equation}
g_{\mu\nu}=h_{ab}e_{\mu}^{a}e_{\nu}^{b}+n_{\mu}n_{\nu}.
\end{equation}
The inverse to the the metric $h_{ab}$ is given by
\begin{equation}
h^{ab}=g^{\mu\nu}e_{\mu}^{a}e_{\nu}^{b}%
\end{equation}
and fulfills the relation
\begin{equation}
g^{\mu\nu}=h^{ab}e_{a}^{\mu}e_{b}^{\nu}+n^{\mu}n^{\nu}%
\end{equation}
where $g^{\mu\nu}$ is the inverse of $g_{\mu\nu}.$ This shows that any tensor
can be projected into the hypersurface using the completeness relations for
the basis $\left\{  e_{\mu}^{a},n_{\mu}\right\}  .$ We now define the Clifford
algebra
\begin{equation}
\left\{  \gamma^{\mu},\gamma^{\nu}\right\}  =-2g^{\mu\nu},\qquad\mu
,\nu=1,\cdots,\mathrm{dim}\,M
\end{equation}
and project these \ to define
\begin{equation}
\gamma^{n}=\gamma^{\mu}n_{\mu},\qquad\gamma^{a}=\gamma^{\mu}e_{\mu}^{a}%
\end{equation}
which satisfy the properties
\begin{equation}
\gamma^{n}\gamma^{n}=-1,\quad\left\{  \gamma^{a},\gamma^{b}\right\}
=-2h^{ab},\quad\left\{  \gamma^{a},\gamma^{n}\right\}  =0
\end{equation}
which follow from the relation
\begin{equation}
\gamma^{\mu}=e_{a}^{\mu}\gamma^{a}+n^{\mu}\gamma^{n}\,.
\end{equation}
We will specialize to manifolds of dimension $4$ so that a local coordinate
system on $\partial M$ will be denoted by $\left\{  y^{a}\right\}  =\left\{
y^{1},y^{2},y^{3}\right\}  $ and for $M$ denoted by $\left\{  x^{\mu}\right\}
=\left\{  x^{1},x^{2},x^{3},x^{4}\right\}  .$ We then define on $\partial M$
\begin{equation}
\chi=-\frac{\sqrt{h}}{3!}\epsilon^{abc}\gamma_{a}\gamma_{b}\gamma_{c}%
,\quad\gamma_{5}=\chi\gamma_{n}%
\end{equation}
which satisfy
\begin{align}
\chi^{2}  &  =1,\qquad\chi\gamma_{a}=\gamma_{a}\chi,\qquad\chi\gamma
_{n}=-\gamma_{n}\chi\\
\gamma_{5}^{2}  &  =1,\qquad\chi\gamma_{5}=-\gamma_{5}\chi.
\end{align}
The normal vector $n^{\mu}$ satisfies the properties (\cite{Poisson} Chapter
3)
\begin{equation}
n_{\mu;\nu}=-K_{ab}e_{\mu}^{a}e_{\nu}^{b}%
\end{equation}
where the covariant derivative $;\nu$ is the space-time covariant derivative
and $K_{ab}$ is the extrinsic curvature whose symmetry follows from the
relation $e_{a;b}^{\mu}=e_{b;a}^{\mu}.$ The Gauss-Weingarten equation is
\cite{Poisson}
\begin{equation}
e_{a;b}^{\mu}=K_{ab}n^{\mu}+^{\left(  3\right)  }\Gamma_{ab}^{c}e_{c}^{\mu}%
\end{equation}
where $^{\left(  3\right)  }\Gamma_{ab}^{c}$ is the three dimensional affine
connection and is given by
\begin{equation}
^{\left(  3\right)  }\Gamma_{ab}^{c}=e_{\mu}^{c}e_{a;\nu}^{\mu}e_{b}^{\nu}.
\end{equation}

\section{Spectral action for noncommutative spaces with boundary}

To compute the spectral action including boundary terms, for noncommutative
spaces, we will utilize the known results which lists the Seeley-deWitt
coefficients for elliptic operators which are the square of the Dirac
operator. An important ingredient in the calculation is to specify the
boundary conditions that must be imposed on the Dirac operator \cite{BG1}
\cite{BG2}. We start with the observation that the Dirac operator must satisfy
the hermiticity condition%
\begin{equation}
\left\langle \Psi,D\Psi\right\rangle =\left\langle D\Psi,\Psi\right\rangle .
\end{equation}
This condition is satisfied provided that the following Dirichlet boundary
condition is imposed (\cite{Vass3} (3.30) p.297)
\begin{equation}
\Pi_{-}\Psi|_{\partial M}=0 \label{boundaryd}%
\end{equation}
where the projector $\Pi_{-}$ is given by
\begin{equation}
\Pi_{-}=\frac{1}{2}\left(  1-\chi\right)  \,.
\end{equation}
We first write the square of the Dirac operator in the form
\begin{align}
P  &  =D^{2}=-\left(  g^{\mu\nu}\partial_{\mu}\partial_{\nu}+\mathbb{A}^{\mu
}+\mathbb{B}\right) \\
&  =-\left(  g^{\mu\nu}\nabla_{\mu}^{^{\prime}}\nabla_{\nu}^{^{\prime}%
}+E\right)
\end{align}
where%
\begin{equation}
\nabla_{\mu}^{^{\prime}}=\partial_{\mu}+\omega_{\mu}^{^{\prime}}%
\end{equation}
and%
\begin{align}
E  &  =\mathbb{B}-g^{\mu\nu}\left(  \partial_{\mu}\omega_{\nu}^{^{\prime}%
}+\omega_{\mu}^{^{\prime}}\omega_{\nu}^{^{\prime}}-\Gamma_{\mu\nu}^{\rho
}\omega_{\rho}^{^{\prime}}\right) \\
\omega_{\mu}^{^{\prime}}  &  =\frac{1}{2}g_{\mu\nu}\left(  \mathbb{A}^{\nu
}+g^{\rho\sigma}\Gamma_{\rho\sigma}^{\nu}\left(  g\right)  \right) \\
\Omega_{\mu\nu}  &  =\partial_{\mu}\omega_{\nu}^{^{\prime}}-\partial_{\nu
}\omega_{\mu}^{^{\prime}}+\omega_{\mu}^{^{\prime}}\omega_{\nu}^{^{\prime}%
}-\omega_{\nu}^{^{\prime}}\omega_{\mu}^{^{\prime}}.
\end{align}
It is convenient to write the Dirac operator in the form
\begin{equation}
D=\gamma^{\mu}\nabla_{\mu}-\Phi
\end{equation}
where $\nabla_{\mu}=\partial_{\mu}+\omega_{\mu}$ with
\begin{equation}
\omega_{\mu}=\frac{1}{4}\omega_{\mu}^{\;\alpha\beta}\gamma_{\alpha\beta}%
\end{equation}
is the spin connection determined by the vanishing of the vierbein covariant
derivative
\begin{equation}
\partial_{\mu}e_{\nu}^{\alpha}-\omega_{\mu}^{\;\alpha\beta}e_{\nu\beta}%
-\Gamma_{\mu\nu}^{\rho}\left(  g\right)  e_{\rho}^{\alpha}=0
\end{equation}
where
\begin{equation}
\Gamma_{\mu\nu}^{\rho}\left(  g\right)  =\frac{1}{2}g^{\rho\sigma}\left(
\partial_{\mu}g_{\sigma\nu}+\partial_{\nu}g_{\mu\sigma}-\partial_{\sigma
}g_{\mu\nu}\right)
\end{equation}
is the Christoffel connection of $g_{\mu\nu}=e_{\mu}^{\alpha}e_{\nu\alpha}.$
Note that $e_{\mu}^{a}$ should not be confused with $e_{\mu}^{\alpha}$ as the
index $\alpha$ refers to the tangent space $T(M)$ and is four dimensional and
has the flat metric $\delta_{\alpha\beta}.$ The covariant derivative
$\nabla_{n}^{^{\prime}}$ is along the normal direction and is defined by
\begin{equation}
n^{\mu}\nabla_{\mu}^{^{\prime}}%
\end{equation}
and the index $n$ always refers to the projection of the vector index along
the normal direction. The boundary conditions for $D^{2}$ are then equivalent
to \cite{BG1}, \cite{BG2}
\begin{equation}
\mathcal{B}_{\chi}\Psi=\Pi_{-}\left(  \Psi\right)  |_{\partial M}\oplus\Pi
_{+}\left(  \nabla_{n}^{^{\prime}}+S\right)  \Pi_{+}\left(  \Psi\right)
|_{\partial M}=0.
\end{equation}
Here $\Pi_{+}=1-\Pi_{-}$, and the operator $S$ \ is
\begin{equation}
S=\Pi_{+}\left(  \gamma_{n}\Phi-\frac{1}{2}\gamma_{n}\gamma^{a}\nabla
_{a}^{^{\prime}}\chi\right)  \Pi_{+}%
\end{equation}
with
\begin{equation}
\nabla_{a}^{^{\prime}}\chi=\partial_{a}\chi+\left[  \omega_{a}^{^{\prime}%
},\chi\right]  =K_{ab}\chi\gamma^{n}\gamma^{b}+\left[  \theta_{a},\chi\right]
\end{equation}
where
\begin{equation}
\theta_{a}=\omega_{a}^{\prime}-\omega_{a}.
\end{equation}
To prove the above relation we write
\begin{equation}
\Pi_{-}\left(  \gamma^{n}\nabla_{n}^{^{\prime}}+\gamma^{a}\nabla_{a}%
^{^{\prime}}-\Phi\right)  \Psi|_{\partial M}=\gamma^{n}\left(  \nabla
_{n}^{^{\prime}}+\gamma_{n}\Phi\right)  \Pi_{+}\Psi|_{\partial M}+\left[
\Pi_{-},\gamma^{a}\nabla_{a}^{^{\prime}}\right]  \Psi|_{\partial M}%
\end{equation}
where we have used $\Pi_{-}\Psi|_{\partial M}=0$ and $\gamma^{a}\nabla
_{a}^{^{\prime}}\left(  \Pi_{-}\Psi|_{\partial M}\right)  =0.$ We then have
\begin{align*}
\left[  \Pi_{-},\gamma^{a}\nabla_{a}^{^{\prime}}\right]  \Psi|_{\partial M}
&  =\frac{1}{2}\gamma^{a}\nabla_{a}^{^{\prime}}\chi\left(  \Pi_{-}\Psi+\Pi
_{+}\Psi\right)  |_{\partial M}\\
&  =\Pi_{-}\left(  \frac{1}{2}\gamma^{a}\nabla_{a}^{^{\prime}}\chi\right)
\Pi_{+}\Psi|_{\partial M}\\
&  =\gamma^{n}\Pi_{+}\left(  \frac{1}{2}\gamma_{n}\gamma^{a}\nabla
_{a}^{^{\prime}}\chi\right)  \Pi_{+}\Psi|_{\partial M}.
\end{align*}
We also have the relations
\begin{align}
E  &  =\gamma^{\mu}\nabla_{\mu}\Phi-\Phi^{2}-\frac{1}{2}\gamma^{\mu\nu}%
\Omega_{\mu\nu},\\
\Omega_{\mu\nu}  &  =\partial_{\mu}\omega_{\nu}^{^{\prime}}-\partial_{\nu
}\omega_{\mu}^{^{\prime}}+\omega_{\mu}^{^{\prime}}\omega_{\nu}^{^{\prime}%
}-\omega_{\nu}^{^{\prime}}\omega_{\mu}^{^{\prime}}.
\end{align}

The Seeley-deWitt coefficients for second order operators on manifolds with
boundary were calculated by Branson and Gilkey \cite{BG1}, \cite{BG2} and are
given by
\begin{equation}
a_{0}\left(  P,\chi\right)  =\frac{1}{16\pi^{2}}%
%TCIMACRO{\dint \limits_{M}}%
%BeginExpansion
{\displaystyle\int\limits_{M}}
%EndExpansion
d^{4}x\sqrt{g}\text{\textrm{Tr}}\left(  1\right)
\end{equation}%
\begin{equation}
a_{1}\left(  P,\chi\right)  =0
\end{equation}%
\begin{equation}
a_{2}\left(  P,\chi\right)  =\frac{1}{96\pi^{2}}\left(
%TCIMACRO{\dint \limits_{M}}%
%BeginExpansion
{\displaystyle\int\limits_{M}}
%EndExpansion
d^{4}x\sqrt{g}\text{\textrm{Tr}}\left(  6E+R\right)  +%
%TCIMACRO{\dint \limits_{\partial M}}%
%BeginExpansion
{\displaystyle\int\limits_{\partial M}}
%EndExpansion
d^{3}x\sqrt{h}\text{\textrm{Tr}}\left(  2K+12S\right)  \right)
\end{equation}%
\begin{align}
a_{3}\left(  P,\chi\right)   &  =\frac{1}{384(4\pi)^{\frac{3}{2}}}%
%TCIMACRO{\dint \limits_{\partial M}}%
%BeginExpansion
{\displaystyle\int\limits_{\partial M}}
%EndExpansion
d^{3}x\sqrt{h}\text{\textrm{Tr}}\left(  96\chi E+3K^{2}+6K_{ab}K^{ab}\right.
\nonumber\\
&  \qquad\qquad\qquad\left.  +96SK+192S^{2}-12\nabla_{a}^{^{\prime}}\chi
\nabla^{^{^{\prime}a}}\chi\right)
\end{align}%
\begin{align}
a_{4}\left(  P,\chi\right)   &  =\frac{1}{360}\frac{1}{16\pi^{2}}\left\{
%TCIMACRO{\dint \limits_{M}}%
%BeginExpansion
{\displaystyle\int\limits_{M}}
%EndExpansion
d^{4}x\sqrt{g}\text{\textrm{Tr}}\left(  60RE+180E^{2}+30\Omega_{\mu\nu}%
\Omega^{\mu\nu}+12\left(  R+5E\right)  _{;\mu}^{\,\,\mu}\right.  \right.
\nonumber\\
&  \hspace{1.1in}\left.  +5R^{2}-2R_{\mu\nu}R^{\mu\nu}+2R_{\mu\nu\rho\sigma
}R^{\mu\nu\rho\sigma}\right) \nonumber\\
&  \hspace{0.7in}+%
%TCIMACRO{\dint \limits_{\partial M}}%
%BeginExpansion
{\displaystyle\int\limits_{\partial M}}
%EndExpansion
d^{3}x\sqrt{h}\text{\textrm{Tr}}\left(  180\chi\nabla_{n}^{^{\prime}%
}E+120EK+20RK\right. \nonumber\\
&  \qquad\qquad\qquad\qquad+4R_{\;nan}^{a}K-12R_{\;nbn}^{a}K_{a}%
^{\;b}+4R_{\;acb}^{c}K^{ab}\nonumber\\
&  \qquad\qquad+\frac{1}{21}\left(  160K^{3}-48KK_{ab}K^{ab}+272K_{\;b}%
^{a}K_{\;c}^{b}K_{\;a}^{c}\right) \nonumber\\
&  \qquad\qquad+720SE+120SR+144SK^{2}+48SK_{ab}K^{ab}+480S^{2}K+480S^{3}%
\nonumber\\
&  \qquad\qquad\left.  \left.  +60\chi\nabla^{^{\prime a}}\chi\Omega
_{an}-12\nabla_{a}^{^{\prime}}\chi\nabla^{^{\prime a}}\chi\left(
K+10S\right)  -24\nabla_{a}^{^{\prime}}\chi\nabla_{b}^{^{\prime}}\chi
K^{ab}\right)  \right\}
\end{align}
The Riemann tensor is defined by
\begin{equation}
R_{\mu\nu\rho\sigma}=g_{\sigma\tau}\left(  \partial_{\mu}\Gamma_{\nu\rho
}^{\tau}-\partial_{\nu}\Gamma_{\mu\rho}^{\tau}+\Gamma_{\mu\kappa}^{\tau}%
\Gamma_{\nu\rho}^{\kappa}-\Gamma_{\nu\kappa}^{\tau}\Gamma_{\mu\rho}^{\kappa
}\right)
\end{equation}
and its contractions are
\begin{equation}
R_{\mu\nu}=g^{\rho\sigma}R_{\mu\rho\sigma\nu},\qquad R=g^{\mu\nu}R_{\mu\nu}%
\end{equation}
We are using the conventions of Gilkey. They are related to the ones used by
Misner-Thorn-Wheeler by
\begin{equation}
R_{\mu\nu\rho\sigma}^{\text{G}}=-R_{\mu\nu\rho\sigma}^{\text{MTW}},\qquad
R_{\mu\nu}^{\text{G}}=R_{\mu\nu}^{\text{MTW}},\text{\qquad}R^{\text{G}%
}=R^{\text{MTW}}%
\end{equation}
The curvature defined by the spin connection is
\begin{equation}
R_{\mu\nu}^{\alpha\beta}\left(  \omega\right)  =\partial_{\mu}\omega_{\nu
}^{\;\alpha\beta}-\partial_{\nu}\omega_{\mu}^{\;\alpha\beta}-\omega_{\mu
}^{\;\alpha\gamma}\omega_{\nu\gamma}^{\;\;\beta}+\omega_{\nu}^{\;\alpha\gamma
}\omega_{\mu\gamma}^{\;\;\beta}%
\end{equation}
and is related to the curvature of the Christoffel connection by
\begin{equation}
R_{\mu\nu}^{\alpha\beta}\left(  \omega\right)  e_{\rho\alpha}e_{\sigma\beta
}=R_{\mu\nu\rho\sigma}^{\text{G}},\qquad R_{\mu\nu}^{\alpha\beta}\left(
\omega\right)  e_{\alpha}^{\mu}e_{\beta}^{\nu}=-R^{\text{G}}.
\end{equation}
We note that the $R$ we used in \cite{ACAC} has the opposite sign to
$R^{\text{G}}$ where the curvature is positive for spheres.

The formulas expressing the projections of the Riemann tensor on the boundary
in terms of the three curvature and the extrinsic curvature are
\begin{align*}
R_{abcd} &  =^{\left(  3\right)  }R_{abcd}+\left(  K_{ac}K_{bd}-K_{ad}%
K_{bc}\right)  \\
R_{nan}^{\quad\,\,a} &  =g^{\mu\rho}n^{\nu}\left(  n_{\mu;\nu\rho}-n_{\mu
;\rho\nu}\right)  =K^{2}-K_{ab}K^{ab}+\text{\textrm{cov. div.}}%
\end{align*}
In particular, we can apply these results to the square of the Dirac operator.
We shall start with the simplest example of the Dirac operator of a pure
gravitational fields, and later generalize the results to the general case of
the standard model.

\section{Spectral action for Riemannian manifolds with boundary}

In this case we have
\begin{equation}
D=\gamma^{\mu}\left(  \partial_{\mu}+\omega_{\mu}\right)  .
\end{equation}
To use the above formulas we have
\begin{equation}
\omega_{\mu}^{^{\prime}}=\omega_{\mu},\quad\Phi=0
\end{equation}
and
\begin{equation}
S=\Pi_{+}\left(  -\frac{1}{2}\gamma_{n}\gamma^{a}\chi K_{ab}\gamma^{n}%
\gamma^{b}\right)  =-\frac{1}{2}K\Pi_{+}%
\end{equation}
where we used $\gamma_{n}\gamma^{a}\chi\gamma^{n}=-\chi\gamma_{n}\gamma
^{a}\gamma^{n}=-\chi\gamma^{a}.$ We also have
\begin{equation}
E=-\frac{1}{4}R,\quad\nabla_{a}^{^{\prime}}\chi=K_{ab}\chi\gamma^{n}\gamma
^{b}.
\end{equation}
Substituting \textrm{Tr}$\left(  1\right)  =4$ and \textrm{Tr}$\left(
S\right)  =-K$ we have
\begin{equation}
a_{0}\left(  P,\chi\right)  =\frac{1}{4\pi^{2}}%
%TCIMACRO{\dint \limits_{M}}%
%BeginExpansion
{\displaystyle\int\limits_{M}}
%EndExpansion
d^{4}x\sqrt{g}.
\end{equation}
Next we calculate
\begin{equation}
a_{2}\left(  P,\chi\right)  =\frac{1}{96\pi^{2}}\left(
%TCIMACRO{\dint \limits_{M}}%
%BeginExpansion
{\displaystyle\int\limits_{M}}
%EndExpansion
d^{4}x\sqrt{g}\text{\textrm{Tr}}\left(  6E+R\right)  +%
%TCIMACRO{\dint \limits_{\partial M}}%
%BeginExpansion
{\displaystyle\int\limits_{\partial M}}
%EndExpansion
d^{3}x\sqrt{h}\text{\textrm{Tr}}\left(  2K+12S\right)  \right)
\end{equation}
we use
\begin{align}
\mathrm{Tr}\left(  6E+R\right)   &  =-\frac{R}{2}\mathrm{Tr}\left(  1\right)
\nonumber\\
\mathrm{Tr}\left(  2K+12S\right)   &  =-K\mathrm{Tr}\left(  1\right)
\end{align}
because Tr$\left(  \Pi_{+}\right)  =\frac{1}{2}.$ In this case Tr$\left(
1\right)  =4$ (trace over Dirac matrices). Substituting into the formula for
$a_{2}$ gives
\begin{equation}
a_{2}\left(  P,\chi\right)  =\frac{1}{24\pi^{2}}\left(
%TCIMACRO{\dint \limits_{M}}%
%BeginExpansion
{\displaystyle\int\limits_{M}}
%EndExpansion
d^{4}x\sqrt{g}\left(  -\frac{1}{2}R\right)  +%
%TCIMACRO{\dint \limits_{\partial M}}%
%BeginExpansion
{\displaystyle\int\limits_{\partial M}}
%EndExpansion
d^{3}x\sqrt{h}\left(  -K\right)  \right)  .
\end{equation}
The important point in the above result is the emergence of the combination
\begin{equation}
-%
%TCIMACRO{\dint \limits_{M}}%
%BeginExpansion
{\displaystyle\int\limits_{M}}
%EndExpansion
d^{4}x\sqrt{g}R-2%
%TCIMACRO{\dint \limits_{\partial M}}%
%BeginExpansion
{\displaystyle\int\limits_{\partial M}}
%EndExpansion
d^{3}x\sqrt{h}K
\end{equation}
as the lowest term of the gravitational action which is known to be the
required correction to the Einstein action including the surface term which
makes the Hamiltonian formalism consistent. The consistency of the variation
of this action is summarized in appendix 2. This is remarkable because both
the sign and the coefficients are correct. The only assumption we made is that
the boundary conditions are taken to satisfy the hermiticity of the Dirac
operator. This is yet another miracle concerning the correct signs obtained in
the spectral action of the Dirac operator. We also notice that the relative
coefficient between $R$ and $K$ depend on the nature of the Laplacian. The
desired answer is obtained naturally for the Dirac operator, but not for a
general Laplacian.

We continue to compute%
\begin{align}
a_{3}\left(  P,\chi\right)   &  =\frac{1}{384(4\pi)^{\frac{3}{2}}}%
%TCIMACRO{\dint \limits_{\partial M}}%
%BeginExpansion
{\displaystyle\int\limits_{\partial M}}
%EndExpansion
d^{3}x\sqrt{h}\mathrm{Tr}\left(  96\chi E+3K^{2}+6K_{ab}K^{ab}\right.
\nonumber\\
&  \qquad\qquad\qquad\left.  +96SK+192S^{2}-12\nabla_{a}^{^{\prime}}\chi
\nabla^{^{^{\prime}a}}\chi\right)  .
\end{align}
We first note that \textrm{Tr}$\left(  96\chi E\right)  =0$ and%
\begin{equation}
\text{\textrm{Tr}}\left(  3K^{2}+6K_{ab}K^{ab}+96SK+192S^{2}\right)
=\text{\textrm{Tr}}\left(  1\right)  \left(  3K^{2}+6K_{ab}K^{ab}\right)
\end{equation}
while
\begin{equation}
\text{\textrm{Tr}}\left(  -12\nabla_{a}^{^{\prime}}\chi\nabla^{^{^{\prime}a}%
}\chi\right)  =\mathrm{Tr}\left(  -12K_{ab}\chi\gamma^{n}\gamma^{b}K_{ac}%
\chi\gamma^{n}\gamma^{c}\right)  =-12K_{ab}K^{ab}\mathrm{Tr}\left(  1\right)
\end{equation}
where we \ have \ used $\chi\gamma^{n}\gamma^{b}\chi\gamma^{n}\gamma
^{c}=-\gamma^{n}\gamma^{b}\gamma^{n}\gamma^{c}=-\gamma^{b}\gamma^{c}.$
Collecting the above terms give
\begin{equation}
a_{3}\left(  P,\chi\right)  =\frac{1}{32(4\pi)^{\frac{3}{2}}}%
%TCIMACRO{\dint \limits_{\partial M}}%
%BeginExpansion
{\displaystyle\int\limits_{\partial M}}
%EndExpansion
d^{3}x\sqrt{h}\left(  K^{2}-2K_{ab}K^{ab}\right)  .
\end{equation}
Finally we turn our attention to the computation of $a_{4}$ which is rather
complicated. First we evaluate%
\begin{align}
&  \text{\textrm{Tr}}\left(  60RE+180E^{2}+30\Omega_{\mu\nu}\Omega^{\mu\nu
}+5R^{2}-2R_{\mu\nu}R^{\mu\nu}+2R_{\mu\nu\rho\sigma}R^{\mu\nu\rho\sigma
}-3R_{;\mu}^{\,\,\mu}\right) \nonumber\\
&  =\text{\textrm{Tr}}\left(  1\right)  \frac{1}{4}\left(  5R^{2}-8R_{\mu\nu
}R^{\mu\nu}-7R_{\mu\nu\rho\sigma}R^{\mu\nu\rho\sigma}-12R_{;\mu}^{\,\,\mu
}\right)  .
\end{align}
We then use the identities
\begin{align}
R_{\mu\nu\rho\sigma}R^{\mu\nu\rho\sigma}-4R_{\mu\nu}R^{\mu\nu}  &  =R^{\ast
}R^{\ast}-R^{2}\\
R_{\mu\nu\rho\sigma}R^{\mu\nu\rho\sigma}-2R_{\mu\nu}R^{\mu\nu}  &  =C_{\mu
\nu\rho\sigma}^{2}-\frac{1}{3}R^{2}%
\end{align}
where $R^{\ast}R^{\ast}=\frac{1}{4}\epsilon^{\mu\nu\rho\sigma}\epsilon
_{\alpha\beta\gamma\delta}R_{\mu\nu}^{\quad\alpha\beta}R_{\rho\sigma}%
^{\quad\gamma\delta}.$ These identities are solved to give
\begin{align}
R_{\mu\nu\rho\sigma}R^{\mu\nu\rho\sigma}  &  =2C_{\mu\nu\rho\sigma}^{2}%
+\frac{1}{3}R^{2}-R^{\ast}R^{\ast}\\
R_{\mu\nu}R^{\mu\nu}  &  =\frac{1}{2}C_{\mu\nu\rho\sigma}^{2}+\frac{1}{3}%
R^{2}-\frac{1}{2}R^{\ast}R^{\ast}%
\end{align}
and can be combined to show that
\begin{equation}
5R^{2}-8R_{\mu\nu}^{2}-7R_{\mu\nu\rho\sigma}^{2}=11R^{\ast}R^{\ast}%
-18C_{\mu\nu\rho\sigma}^{2}.
\end{equation}
We continue by evaluating
\begin{align*}
&  \text{\textrm{Tr}}\left(  180\chi\nabla_{n}^{^{\prime}}%
E+120EK+20RK+4R_{\;nan}^{a}K-12R_{\;nbn}^{a}K_{a}^{\;b}+4R_{\;acb}^{c}%
K^{ab}\right) \\
&  =\text{\textrm{Tr}}\left(  1\right)  \left(  -10RK+4R_{\;nan}%
^{a}K-12R_{\;nbn}^{a}K_{a}^{\;b}+4R_{\;acb}^{c}K^{ab}\right)
\end{align*}
In addition the expression%
\[
\frac{1}{21}\mathrm{Tr}\left(  160K^{3}-48KK_{ab}K^{ab}+272K_{\;b}^{a}%
K_{\;c}^{b}K_{\;a}^{c}\right)
\]
cannot be simplified. Next we have
\begin{align*}
&  \text{\textrm{Tr}}\left(  720SE+120SR+144SK^{2}+48SK_{ab}K^{ab}%
+480S^{2}K+480S^{3}\right. \\
&  \ \left.  +60\chi\nabla^{^{\prime a}}\chi\Omega_{an}-12\nabla_{a}%
^{^{\prime}}\chi\nabla^{^{\prime a}}\chi\left(  K+10S\right)  -24\nabla
_{a}^{^{\prime}}\chi\nabla_{b}^{^{\prime}}\chi K^{ab}\right) \\
&  =\text{\textrm{Tr}}\left(  720\left(  -\frac{K}{4}\right)  \left(
-\frac{R}{4}\right)  +120\left(  -\frac{K}{4}R\right)  +144\left(  -\frac
{K}{4}K^{2}\right)  +48\left(  -\frac{K}{4}K_{ab}K^{ab}\right)  \right. \\
&  +480\frac{K^{2}}{4}\frac{1}{2}K+480\left(  -\frac{K^{3}}{8}\frac{1}%
{2}\right)  +60K_{ab}\chi^{2}\gamma^{n}\gamma^{b}\frac{1}{4}R_{an}%
^{\quad\alpha\beta}\gamma_{\alpha\beta}\\
&  \left.  -12\left(  K-\frac{10}{4}K\right)  K_{ac}\chi\gamma^{n}\gamma
^{c}K_{ad}\chi\gamma^{n}\gamma^{d}-24K_{ac}\chi\gamma^{n}\gamma^{c}K_{bd}%
\chi\gamma^{n}\gamma^{d}K^{ab}\right)  .
\end{align*}
The factors of $\frac{1}{2}$ appearing above are due to the presence of
$\ \Pi_{+}=\frac{1}{2}\left(  1+\chi\right)  .$ Evaluating the traces, the
above expression simplifies to%

\[
\text{\textrm{Tr}}\left(  1\right)  \left(  15KR-6K^{3}+30R_{an}^{\quad
bn}K^{ab}+6KK_{ab}K^{ab}-24K_{a}^{b}K_{b}^{c}K_{c}^{a}\right)
\]
Combining all the above terms give%

\begin{align}
&  \text{\textrm{Tr}}\left(  1\right)  \left(  5RK+4R_{\;nan}^{a}%
K+18R_{\;nbn}^{a}K_{a}^{\;b}+4R_{\;acb}^{c}K^{ab}\right. \nonumber\\
&  \qquad\qquad\left.  +\frac{2}{21}\left(  17K^{3}+39KK_{ab}K^{ab}%
-116K_{\;b}^{a}K_{\;c}^{b}K_{\;a}^{c}\right)  \right)  .
\end{align}
Thus the final expression for $a_{4}$ is given by
\begin{align}
a_{4}\left(  P,\chi\right)   &  =\frac{1}{360}\frac{1}{16\pi^{2}}\left\{
%TCIMACRO{\dint \limits_{M}}%
%BeginExpansion
{\displaystyle\int\limits_{M}}
%EndExpansion
d^{4}x\sqrt{g}\left(  5R^{2}-8R_{\mu\nu}^{2}-7R_{\mu\nu\rho\sigma}%
^{2}-12R_{;\mu}^{\,\,\mu}\right)  \right. \nonumber\\
&  +4%
%TCIMACRO{\dint \limits_{\partial M}}%
%BeginExpansion
{\displaystyle\int\limits_{\partial M}}
%EndExpansion
d^{3}x\sqrt{h}\left(  \frac{2}{21}\left(  17K^{3}+39KK_{ab}K^{ab}%
-116K_{a}^{\;b}K_{b}^{\;c}K_{c}^{\;a}\right)  \right. \nonumber\\
&  \hspace{0.6in}\left.  +\left(  5RK+4KR_{\;nan}^{a}+4K_{ab}R_{\;acb}%
^{c}+18R_{anbn}K^{ab}\right)  \right\}  .
\end{align}
Collecting terms, the spectral action is then given by
\begin{equation}
I=2\left(  f_{4}\Lambda^{4}a_{0}+f_{2}\Lambda^{2}a_{2}+f_{1}\Lambda
a_{3}+f_{0}a_{4}\right)  +O\left(  \frac{1}{\Lambda^{2}}\right)  .
\end{equation}

\section{Spectral Action for the noncommutative space of the Standard Model
with boundary}

It was shown recently \cite{Beggar}, \cite{Why}\ by making the basic
assumption at some high-energy scale that space-time is described by a
noncommutative space which is a product of a continuous four-dimensional
Riemannian manifold times a finite space, it is possible to almost uniquely
determine the algebra and Hilbert space of the finite space. The main
constraints come from the axioms of noncommutative geometry, as well as from
the physical requirement that there is a mixing between the fermions and their
conjugates, which turns out to imply that the neutrinos get a Majorana mass
through the see-saw mechanism. Under these conditions, the algebra is given by
$\mathcal{A}=C^{\infty}\left(  M\right)  \otimes\mathcal{A}_{F}$ where the
algebra $\mathcal{A}_{F}$ is finite dimensional, $\mathcal{A}_{F}%
=\mathbb{C}\oplus\mathbb{H}\oplus M_{3}\left(  \mathbb{C}\right)  ,$ and
$\mathbb{H}\subset M_{2}\left(  \mathbb{C}\right)  $ is the algebra of
quaternions. The important point to emphasize is that the number of fermions
is predicted to be $4^{2}=16$, and the representations of the fermions follow
from the decomposition of the representation $\left(  4,4\right)  $ with
respect to the subalgebra $\mathbb{C}\oplus\mathbb{H}\oplus M_{3}\left(
\mathbb{C}\right)  $ of $\mathbb{H}\oplus\mathbb{H}\oplus M_{4}\left(
\mathbb{C}\right)  .$ The spectral geometry of $\mathcal{A}$ is given by the
product rule
\[
\mathcal{H}=L^{2}\left(  M,S\right)  \otimes\mathcal{H}_{F},\quad
D=D_{M}\otimes1+\gamma_{5}\otimes D_{F}%
\]
where $L^{2}\left(  M,S\right)  $ is the Hilbert space of $L^{2}$ spinors, and
$D_{M}$ is the Dirac operator of the Levi-Civita spin connection on $M.$ The
operator $D_{F}$ anticommutes with the chirality operator $\gamma_{F}$ on
$\mathcal{H}_{F}.$ The spectral geometry does not change if one replaces $D$
\ by the equivalent operator
\[
D=D_{M}\otimes\gamma_{F}+1\otimes D_{F}%
\]
but this equivalence fails when $M$ \ has a boundary and it is only the latter
choice which has conceptual meaning since $\gamma_{5}$ no longer anticommutes
with $D_{M}$ when $\partial M\neq\emptyset$. The noncommutative space defined
by a spectral triple has to satisfy the basic axioms of noncommutative
geometry. The charge conjugation operator $J$ \ for the product geometry is
then given by
\[
J=J_{M}\,\,\gamma_{5}\otimes J_{F}%
\]
which commutes with the operator $D$ \ since in even dimension $J_{M}$
commutes with $D_{M}$ while in dimension $6$ modulo $8$, $J_{F}$ anticommutes
with $\gamma_{F}$. The $KO$-dimension of the noncommutative space must be
taken to be equal to $6$ to insure that the fermions and their conjugates are
not independent, and thus avoiding the fermion doubling problem.

Our main interest now is to derive again the spectral action of the standard
model, including boundary contributions. The computations are very
complicated, and because of this it is important to device a way to make this
calculation tractable. The starting point is the observation that the inner
fluctuations under the action of the unitary transformations of the algebra,
forces the Dirac operator to be modified to
\begin{equation}
D\rightarrow D_{A}=D+A+JAJ^{-1},\quad A=%
%TCIMACRO{\dsum }%
%BeginExpansion
{\displaystyle\sum}
%EndExpansion
a\left[  D,b\right]  .
\end{equation}
The Dirac operator acts on the $96$ dimensional space of the three families of
$16$ dimensional spinors and their conjugates, and splits into a leptonic
sector and a quark sector. It turns out that we can get a handle on this
calculation by considering first the much simpler problem of a Dirac operator
of the type:
\begin{equation}
D_{A}=\left(
\begin{array}
[c]{cc}%
\gamma^{\mu}\left(  \left(  \partial_{\mu}+\omega_{\mu}\right)  1_{N}+B_{\mu
}\right)  \otimes\gamma_{F} & H\\
H^{\dagger} & \gamma^{\mu}\left(  \left(  \partial_{\mu}+\omega_{\mu}\right)
1_{M}+B_{\mu}\right)  \otimes\gamma_{F}%
\end{array}
\right)
\end{equation}
where $B_{\mu}$ is an $N\times N$ matrix valued gauge fields. We shall then
define substitutions which will enable us to find the answer for the general
case without much difficulty.

Having defined $D=\gamma^{\mu}\nabla_{\mu}-\Phi$ we can easily deduce that
\begin{equation}
\Phi=-\left(
\begin{array}
[c]{cc}%
\gamma^{\mu}B_{\mu}\otimes\gamma_{F} & H\\
H^{\dagger} & \gamma^{\mu}B_{\mu}\otimes\gamma_{F}%
\end{array}
\right)  .
\end{equation}
We then evaluate $D^{2}$ and put it in canonical form to find that
\begin{equation}
\mathbb{A}^{\mu}=\left(  2g^{\mu\nu}\omega_{\nu}-g^{\rho\sigma}\Gamma
_{\rho\sigma}^{\mu}\right)  1_{N+2}+2g^{\mu\nu}B_{\nu}1_{2}%
\end{equation}%
\begin{align}
\mathbb{B}  &  =\left(  \partial^{\mu}\omega_{\mu}+\omega^{\mu}\omega_{\mu
}-\Gamma^{\mu}\omega_{\mu}-R\right)  1_{N+2}+2\omega_{\mu}g^{\mu\nu}B_{\nu
}1_{2}\nonumber\\
&  +\left(
\begin{array}
[c]{cc}%
X\otimes\gamma_{F} & \gamma^{\mu}\nabla_{\mu}H\\
-\gamma^{\mu}\nabla_{\mu}H^{\dagger} & X\otimes\gamma_{F}%
\end{array}
\right)
\end{align}
where
\begin{align}
X  &  =\left(  \partial^{\mu}+\omega^{\mu}-\Gamma^{\mu}\right)  B_{\mu}%
-\frac{1}{2}\gamma^{\mu\nu}F_{\mu\nu}+B^{\mu}B_{\mu}\\
\nabla_{\mu}H  &  =\partial_{\mu}H+\left[  B_{\mu},H\right] \\
F_{\mu\nu}\left(  B\right)   &  =\partial_{\mu}B_{\nu}-\partial_{\nu}B_{\mu
}+\left[  B_{\mu},B_{\nu}\right]
\end{align}
From these we can construct $\omega_{\mu}^{^{\prime}},$ $E$ and $\Omega
_{\mu\nu}:$%
\begin{equation}
\omega_{\mu}^{^{\prime}}=\left(  \omega_{\mu}\right)  1_{N+M}+B_{\mu}1_{2}%
\end{equation}%
\begin{equation}
\Omega_{\mu\nu}=\frac{1}{4}\left(  R_{\mu\nu}^{\quad\alpha\beta}\gamma
_{\alpha\beta}\right)  1_{N+2}+F_{\mu\nu}\left(  B\right)  1_{2}%
\end{equation}%
\begin{equation}
E=\left(  -\frac{1}{4}R\right)  1_{N+2}+\left(
\begin{array}
[c]{cc}%
-\frac{1}{2}\gamma^{\mu\nu}F_{\mu\nu}\left(  B\right)  -HH^{\dagger} &
\gamma^{\mu}\nabla_{\mu}H\otimes\gamma_{F}\\
-\gamma^{\mu}\nabla_{\mu}H^{\dagger}\otimes\gamma_{F} & -\frac{1}{2}%
\gamma^{\mu\nu}F_{\mu\nu}\left(  B\right)  -H^{\dagger}H
\end{array}
\right)
\end{equation}%
\begin{equation}
\theta_{\mu}=\omega_{\mu}^{^{\prime}}-\omega_{\mu}=B_{\mu}1_{2}.
\end{equation}
From these relations, and assuming that the boundary condition that the
\ normal components of the vectors vanish on the boundary
\[
B_{n}|_{\partial M}=0
\]
we deduce that:
\begin{equation}
S=\Pi_{+}\left(  -\frac{1}{2}K1_{N+2}\right)  .
\end{equation}
The reason for the vanishing of all Higgs and vector terms from the $S$ \ is
the relation $\Pi_{+}\gamma_{n}\,\Pi_{+}=0$ and $\Pi_{+}\gamma_{n}\gamma
_{a}\,\Pi_{+}=0.$

We now derive the Seeley-de Witt coefficients $a_{n}.$ Starting with $a_{0}$
we have:
\begin{equation}
a_{0}\left(  P,\chi\right)  =\frac{\text{\textrm{Tr}}\left(  1_{N+2}\right)
}{16\pi^{2}}%
%TCIMACRO{\dint \limits_{M}}%
%BeginExpansion
{\displaystyle\int\limits_{M}}
%EndExpansion
d^{4}x\sqrt{g}.
\end{equation}
(To facilitate going to the standard model at a later stage and to make use of
the results there, we will include all numerical factors in the \textrm{Tr}%
(..), thus in what follows we will not take out the factor $4$ coming from
tracing over Dirac matrices). Next we find $a_{2}$ by evaluating the various parts%

\begin{equation}
\text{\textrm{Tr}}\left(  6E+R\right)  =-\frac{R}{2}\text{\textrm{Tr}}\left(
1_{N+2}\right)  -12\text{\textrm{Tr}}H^{\dagger}H
\end{equation}

\begin{equation}
\text{\textrm{Tr}}\left(  2K+12S\right)  =-K\text{\textrm{Tr}}\left(
1\right)
\end{equation}
because Tr$\left(  \Pi_{+}\right)  =\frac{1}{2}.$ Thus\newline%
\begin{equation}
a_{2}\left(  P,\chi\right)  =-\frac{1}{96\pi^{2}}\left(
%TCIMACRO{\dint \limits_{M}}%
%BeginExpansion
{\displaystyle\int\limits_{M}}
%EndExpansion
d^{4}x\sqrt{g}\left(  \frac{1}{2}R\text{\textrm{Tr}}\left(  1_{N+2}\right)
+12\text{\textrm{Tr}}\left(  H^{\dagger}H\right)  \right)  +%
%TCIMACRO{\dint \limits_{\partial M}}%
%BeginExpansion
{\displaystyle\int\limits_{\partial M}}
%EndExpansion
d^{3}x\sqrt{h}\text{\textrm{Tr}}\left(  1_{N+2}\right)  K\right)  .
\end{equation}
Next we find $a_{3}$ by computing its parts:%

\begin{equation}
\text{\textrm{Tr}}\left(  96\chi E\right)  =0
\end{equation}

\begin{equation}
\text{\textrm{Tr}}\left(  3K^{2}+6K_{ab}K^{ab}+96SK+192S^{2}\right)
=3\text{\textrm{Tr}}\left(  1_{N+2}\right)  \left(  K^{2}+2K_{ab}%
K^{ab}\right)
\end{equation}

\begin{equation}
\text{\textrm{Tr}}\left(  -12\nabla_{a}^{^{\prime}}\chi\nabla^{^{^{\prime}a}%
}\chi\right)  =-12K_{ab}K^{ab}\text{\textrm{Tr}}\left(  1_{N+2}\right)  .
\end{equation}
Therefore after substituting, $a_{3}$ simplifies to:%

\begin{equation}
a_{3}\left(  P,\chi\right)  =\frac{1}{128(4\pi)^{\frac{3}{2}}}%
%TCIMACRO{\dint \limits_{\partial M}}%
%BeginExpansion
{\displaystyle\int\limits_{\partial M}}
%EndExpansion
d^{3}x\sqrt{h}\left(  \left(  K^{2}-2K_{ab}K^{ab}\right)  \text{\textrm{Tr}%
}\left(  1_{N+2}\right)  \right)  .
\end{equation}
Finally, we turn our attention to $a_{4}$ and concentrate on the terms which
were not present in the pure Riemannian case. First we simplify the combination:%

\begin{align*}
&  \text{\textrm{Tr}}\left(  60RE+180E^{2}+30\Omega_{\mu\nu}\Omega^{\mu\nu
}+5R^{2}-2R_{\mu\nu}R^{\mu\nu}+2R_{\mu\nu\rho\sigma}R^{\mu\nu\rho\sigma
}+12\left(  R+5E\right)  _{;\mu}^{\,\,\mu}\right) \\
&  =\text{\textrm{Tr}}\left(  \left(  -15+\frac{45}{4}+5\right)  R^{2}+\left(
2-\frac{15}{4}\right)  R_{\mu\nu\rho\sigma}R^{\mu\nu\rho\sigma}-2R_{\mu\nu
}R^{\mu\nu}-3\left(  R\right)  _{;\mu}^{\,\,\mu}-120\left(  H^{\dagger
}H\right)  _{;\mu}^{\,\,\mu}\right) \\
&  \qquad\qquad\left.  +\left(  180-120\right)  RH^{\dagger}H-(180-60)F_{\mu
\nu}^{2}+360\left(  H^{\dagger}H\right)  ^{2}+360\nabla_{\mu}H^{\dagger}%
\nabla_{\mu}H\right) \\
&  =\text{\textrm{Tr}}\left(  1_{N+2}\right)  \frac{1}{4}\left(  -18C_{\mu
\nu\rho\sigma}^{2}+11R^{\ast}R^{\ast}-12\left(  R\right)  _{;\mu}^{\,\,\mu
}\right) \\
&  \qquad\qquad+360\left(  \text{\textrm{Tr}}\left(  H^{\dagger}H\right)
^{2}+\text{\textrm{Tr}}\nabla_{\mu}H^{\dagger}\nabla_{\mu}H+\frac{1}%
{6}R\text{\textrm{Tr}}H^{\dagger}H-\frac{1}{3}\text{\textrm{Tr}}F_{\mu\nu}%
^{2}\right)  -120\text{\textrm{Tr}}\left(  H^{\dagger}H\right)  _{;\mu
}^{\,\,\mu}%
\end{align*}
Next we consider:%

\begin{align*}
&  \text{\textrm{Tr}}\left(  180\chi\nabla_{n}^{^{\prime}}%
E+120EK+20RK+4R_{\;nan}^{a}K-12R_{\;nbn}^{a}K_{a}^{\;b}+4R_{\;acb}^{c}%
K^{ab}\right) \\
&  =\text{\textrm{Tr}}\left(  1_{N+2}\right)  \left(  0+120\left(  -\frac
{R}{4}-2H^{\dagger}H\right)  K+20RK+4R_{\;nan}^{a}K-12R_{\;nbn}^{a}K_{a}%
^{\;b}+4R_{\;acb}^{c}K^{ab}\right) \\
&  =\text{\textrm{Tr}}\left(  1_{N+2}\right)  \left(  -10RK+4R_{\;nan}%
^{a}K-12R_{\;nbn}^{a}K_{a}^{\;b}+4R_{\;acb}^{c}K^{ab}\right)  -240K\,\text{Tr}%
H^{\dagger}H.
\end{align*}
where the only change from the purely gravitational case is the addition of
$-240K\,$Tr$H^{\dagger}H.$ The combination%
\[
\frac{1}{21}\text{\textrm{Tr}}\left(  160K^{3}-48KK_{ab}K^{ab}+272K_{\;b}%
^{a}K_{\;c}^{b}K_{\;a}^{c}\right)
\]
does not simplify. Next we consider
\begin{align*}
&  \text{\textrm{Tr}}\left(  720SE+120SR+144SK^{2}+48SK_{ab}K^{ab}%
+480S^{2}K+480S^{3}\right. \\
&  \ \quad\quad\left.  +60\chi\nabla^{^{\prime a}}\chi\Omega_{an}-12\nabla
_{a}^{^{\prime}}\chi\nabla^{^{\prime a}}\chi\left(  K+10S\right)
-24\nabla_{a}^{^{\prime}}\chi\nabla_{b}^{^{\prime}}\chi K^{ab}\right)
\end{align*}
and note that the only change from the purely gravitational case is that
\textrm{Tr}$\left(  720SE\right)  $ will give the extra term $360K$
\textrm{Tr}$H^{\dagger}H$. \ The next three contributions

\begin{align*}
\mathrm{Tr}\left(  60\chi\nabla^{^{\prime a}}\chi\Omega_{an}\right)   &
=\mathrm{Tr}\left(  60\chi K_{ab}\chi\gamma^{n}\gamma^{b}\left(  \frac{1}%
{4}R_{an}^{\quad\alpha\beta}\gamma_{\alpha\beta}+F_{an}\right)  \right) \\
&  =\text{\textrm{Tr}}\left(  1_{N+2}\right)  \left(  30R_{\;nbn}^{a}%
K_{a}^{\;b}\right)
\end{align*}

\[
\mathrm{Tr}\left(  -12\nabla_{a}^{^{\prime}}\chi\nabla^{^{\prime a}}%
\chi\left(  K+10S\right)  \right)  =\text{\textrm{Tr}}\left(  1_{N+2}\right)
\left(  18KK_{ab}K^{ab}\right)
\]

\[
\mathrm{Tr}\left(  -24\nabla_{a}^{^{\prime}}\chi\nabla_{b}^{^{\prime}}\chi
K^{ab}\right)  =\text{\textrm{Tr}}\left(  1_{N+2}\right)  \left(
-24K_{\;b}^{a}K_{\;c}^{b}K_{\;a}^{c}\right)
\]
remain unchanged. Therefore the total change for the boundary term from the
purely gravitational case is the addition of%

\begin{equation}
\left(  -240+360\right)  K\text{\textrm{Tr}}H^{\dagger}H=120K\text{\textrm{Tr}%
}H^{\dagger}H.
\end{equation}
Combining all the above terms give%

\begin{align}
a_{4}\left(  P,\chi\right)   &  =\frac{1}{16\pi^{2}}\left\{
%TCIMACRO{\dint \limits_{M}}%
%BeginExpansion
{\displaystyle\int\limits_{M}}
%EndExpansion
d^{4}x\sqrt{g}\left\{  \frac{1}{360}\left(  \frac{11}{4}R^{\ast}R^{\ast}%
-\frac{9}{2}C_{\mu\nu\rho\sigma}^{2}-3\left(  R\right)  _{;\mu}^{\,\,\mu
}\right)  \text{\textrm{Tr}}1_{N+2}\right.  \right. \nonumber\\
&  \hspace{0.8in}\left.  +\text{\textrm{Tr}}\left(  \left\vert \nabla_{\mu
}H\right\vert ^{2}+\left(  H^{\dagger}H\right)  ^{2}+\frac{1}{6}RH^{\dagger
}H-\frac{1}{3}F_{\mu\nu}^{\,2}\right)  -\frac{1}{3}\text{\textrm{Tr}}\left(
H^{\dagger}H\right)  _{;\mu}^{\,\,\mu}\right\} \nonumber\\
&  +\frac{1}{360}%
%TCIMACRO{\dint \limits_{\partial M}}%
%BeginExpansion
{\displaystyle\int\limits_{\partial M}}
%EndExpansion
d^{3}x\sqrt{h}\left\{  \frac{2}{21}\left(  17K^{3}+39KK_{ab}K^{ab}%
-116K_{a}^{\;b}K_{b}^{\;c}K_{c}^{\;a}\right)  \text{\textrm{Tr}}1_{N+2}\right.
\nonumber\\
&  \left.  \left.  +\left(  5RK+4KR_{\;nan}^{a}+4K_{ab}R_{\;acb}%
^{c}+18R_{anbn}K^{ab}\right)  \text{\textrm{Tr}}1_{N+2}+120K\text{\textrm{Tr}%
}H^{\dagger}H\right\}  \right\}  .
\end{align}
We can now apply these formulas to the Dirac operator of the Standard model by
making the following substitutions%

\begin{equation}
\mathrm{Tr}1_{N+2}\rightarrow\left(  96+288\right)  =384.
\end{equation}
For the trace in the leptonic sector we have
\[
\mathrm{Tr}(1)=4\cdot3\cdot4\cdot2=96
\]
where the first $4$ is from the trace of gamma matrices, the $3$ is for number
of generations, the $4$ is the dimension of the basis for leptons, and $2$ is
for summing over fermions and conjugate fermions. In the quarks sector we
have
\[
\mathrm{Tr}(1)=4\cdot3\cdot4\cdot2\cdot3=288
\]
where the last factor of $3$ is for color. Next for the Higgs field we make
the substitution%

\begin{equation}
\mathrm{Tr}\left(  H^{\dagger}H\right)  \rightarrow8\left(  a\left\vert
\varphi\right\vert ^{2}+\frac{1}{2}c\right)
\end{equation}
where
\begin{equation}
a=\,\mathrm{tr}\left(  3\left\vert k^{u}\right\vert ^{2}+3\left\vert
k^{d}\right\vert ^{2}+\left\vert k^{e}\right\vert ^{2}+\left\vert k^{\nu
}\right\vert ^{2}\right)  \quad\text{and\quad}c=\,\mathrm{tr}\left(
\left\vert k^{\nu_{R}}\right\vert ^{2}\right)  .
\end{equation}
and the factor $8=4\cdot2$ with the $4$ coming from trace of Dirac gamma
matrices, and $2$ because of summing over fermionic and conjugate fermions.
The term $4c$ appears because of the mass mixing term between the fermions and
their conjugates. Next we have%

\begin{equation}
-\frac{1}{3}\mathrm{Tr}\left(  F_{\mu\nu}^{2}\right)  \rightarrow8\left[
g_{3}^{2}\left(  G_{\mu\nu}^{i}\right)  ^{2}+g_{2}^{2}\left(  F_{\mu\nu
}^{\alpha}\right)  ^{2}+\frac{5}{3}g_{1}^{2}\left(  B_{\mu\nu}\right)
^{2}\right]
\end{equation}
where $G_{\mu\nu}^{i}$, $F_{\mu\nu}^{\alpha}$ and $B_{\mu\nu}$ are the
$SU(3),$ $SU(2)$ and $U(1)$ gauge curvatures. The contributions from the
leptonic sector are
\[
-\frac{1}{16\pi^{2}}\frac{1}{6}\mathrm{Tr}F_{\mu\nu}F^{\mu\nu}=-\frac{1}%
{16\pi^{2}}\frac{24}{6}\left(  2\left(  -i\frac{g_{2}}{2}\right)  ^{2}%
F_{\mu\nu}^{\alpha}F^{\mu\nu\alpha}+2\left(  i\frac{g_{1}}{2}\right)
^{2}B_{\mu\nu}B^{\mu\nu}+\left(  ig_{1}\right)  ^{2}B_{\mu\nu}B^{\mu\nu
}\right)
\]
and from the quarks sector%
\begin{align*}
-\frac{1}{16\pi^{2}}\frac{1}{6}\mathrm{Tr}F_{\mu\nu}F^{\mu\nu}  &  =-\frac
{24}{6}\frac{1}{16\pi^{2}}\left(  2\cdot3\left(  -i\frac{g_{2}}{2}\right)
^{2}F_{\mu\nu}^{\alpha}F^{\mu\nu\alpha}+2\cdot3\left(  -i\frac{g_{1}}%
{6}\right)  ^{2}B_{\mu\nu}B^{\mu\nu}\right. \\
&  \left.  +3\cdot\left(  \frac{i}{3}g_{1}\right)  ^{2}B_{\mu\nu}B^{\mu\nu
}+3\cdot\left(  -\frac{2i}{3}g_{1}\right)  ^{2}B_{\mu\nu}B^{\mu\nu}%
+2\cdot4\cdot\left(  -\frac{i}{2}g_{3}\right)  ^{2}G_{\mu\nu}^{i}G^{\mu\nu
i}\right)
\end{align*}
where $24=4\cdot3\cdot2$, the $4$ is due to trace on gamma matrices, $3$ from
generations and $2$ from fermions and their conjugates. $\ $Next, $\frac
{180}{360}$\textrm{Tr}$\left(  E^{2}\right)  $ gives the extra contribution
of
\begin{equation}
\frac{1}{2}\left(  2e\left\vert \varphi\right\vert ^{2}+\frac{1}{2}d\right)
\end{equation}
where
\begin{equation}
d=\mathrm{tr}\left(  \left\vert k^{\nu_{R}}\right\vert ^{4}\right)
\quad\text{and}\quad e=\mathrm{tr}\left(  \left\vert k^{\nu_{R}}\right\vert
^{2}\left\vert k^{\nu}\right\vert ^{2}\right)  .
\end{equation}
Finally%

\begin{equation}
\mathrm{Tr}\left(  H^{\dagger}H\right)  ^{2}\rightarrow b\left\vert
\varphi\right\vert ^{4}%
\end{equation}
where
\begin{equation}
b=\,\mathrm{tr}\left(  3\left\vert k^{u}\right\vert ^{4}+3\left\vert
k^{d}\right\vert ^{4}+\left\vert k^{e}\right\vert ^{4}+\left\vert k^{\nu
}\right\vert ^{4}\right)  .
\end{equation}
Summarizing, we have
\begin{equation}
a_{0}=\frac{384}{16\pi^{2}}%
%TCIMACRO{\dint \limits_{M}}%
%BeginExpansion
{\displaystyle\int\limits_{M}}
%EndExpansion
d^{4}x\sqrt{g}%
\end{equation}%
\begin{equation}
a_{2}=\frac{4}{\pi^{2}}\left(
%TCIMACRO{\dint \limits_{M}}%
%BeginExpansion
{\displaystyle\int\limits_{M}}
%EndExpansion
d^{4}x\sqrt{g}\left(  -\frac{1}{2}R-\frac{1}{4}\left(  a\left\vert
\varphi\right\vert ^{2}+\frac{1}{2}c\right)  \right)  -%
%TCIMACRO{\dint \limits_{\partial M}}%
%BeginExpansion
{\displaystyle\int\limits_{\partial M}}
%EndExpansion
d^{3}x\sqrt{h}K\right)
\end{equation}%
\begin{equation}
a_{3}=\frac{1}{(4\pi)^{\frac{3}{2}}}%
%TCIMACRO{\dint \limits_{\partial M}}%
%BeginExpansion
{\displaystyle\int\limits_{\partial M}}
%EndExpansion
d^{3}x\sqrt{h}\left(  3\left(  K^{2}-2K_{ab}K^{ab}\right)  \right)
\end{equation}%
\begin{align}
a_{4}  &  =\frac{1}{16\pi^{2}}\left\{
%TCIMACRO{\dint \limits_{M}}%
%BeginExpansion
{\displaystyle\int\limits_{M}}
%EndExpansion
d^{4}x\sqrt{g}\left(  \frac{384}{360}\frac{1}{4}\left(  -18C_{\mu\nu\rho
\sigma}^{2}+11R^{\ast}R^{\ast}-12\left(  R\right)  _{;\mu}^{\,\,\mu}\right)
\right.  \right. \nonumber\\
&  +8\left(  a\left\vert D_{\mu}\varphi\right\vert ^{2}+\frac{1}{6}R\left(
a\left\vert \varphi\right\vert ^{2}+\frac{1}{2}c\right)  +b\left\vert
\varphi\right\vert ^{4}+\frac{1}{2}d-\frac{1}{3}a\left(  \left\vert
\varphi\right\vert ^{2}\right)  _{;\mu}^{\,\,\mu}\right) \nonumber\\
&  \left.  +8\left(  g_{3}^{2}\left(  G_{\mu\nu}^{i}\right)  ^{2}+g_{2}%
^{2}\left(  F_{\mu\nu}^{\alpha}\right)  ^{2}+\frac{5}{3}g_{1}^{2}\left(
B_{\mu\nu}\right)  ^{2}\right)  \right) \nonumber\\
&  +%
%TCIMACRO{\dint \limits_{\partial M}}%
%BeginExpansion
{\displaystyle\int\limits_{\partial M}}
%EndExpansion
d^{3}x\sqrt{h}\left(  \frac{1}{3}K\left(  a\left\vert \varphi\right\vert
^{2}+\frac{1}{2}c\right)  \right. \nonumber\\
&  +\frac{384}{360}\left(  5RK+4KR_{\;nan}^{a}+4K_{ab}R_{\;acb}^{c}%
+18R_{anbn}K^{ab}\right) \nonumber\\
&  \left.  +\frac{384}{360}\frac{2}{21}\left(  17K^{3}+39KK_{ab}%
K^{ab}-116K_{a}^{\;b}K_{b}^{\;c}K_{c}^{\;a}\right)  \right) \nonumber\\
&  =\frac{1}{2\pi^{2}}\left\{
%TCIMACRO{\dint \limits_{M}}%
%BeginExpansion
{\displaystyle\int\limits_{M}}
%EndExpansion
d^{4}x\sqrt{g}\left(  \left(  -\frac{3}{5}C_{\mu\nu\rho\sigma}^{2}+\frac
{11}{30}R^{\ast}R^{\ast}-\frac{2}{5}\left(  R\right)  _{;\mu}^{\,\,\mu
}\right)  \right.  \right. \nonumber\\
&  \left.  +a\left\vert D_{\mu}\varphi\right\vert ^{2}+\frac{1}{6}R\left(
a\left\vert \varphi\right\vert ^{2}+\frac{1}{2}c\right)  +b\left\vert
\varphi\right\vert ^{4}+2e\left\vert \varphi\right\vert ^{2}+\frac{1}%
{2}d-\frac{1}{3}a\left(  \left\vert \varphi\right\vert ^{2}\right)  _{;\mu
}^{\,\,\mu}\right) \nonumber\\
&  +%
%TCIMACRO{\dint \limits_{\partial M}}%
%BeginExpansion
{\displaystyle\int\limits_{\partial M}}
%EndExpansion
d^{3}x\sqrt{h}\left(  \frac{1}{3}K\left(  a\left\vert \varphi\right\vert
^{2}+\frac{1}{2}c\right)  \right. \nonumber\\
&  +\frac{2}{15}\left(  5RK+4KR_{\;nan}^{a}+4K_{ab}R_{\;acb}^{c}%
+18R_{anbn}K^{ab}\right) \nonumber\\
&  \left.  \left.  +\frac{4}{315}\left(  17K^{3}+39KK_{ab}K^{ab}%
-116K_{a}^{\;b}K_{b}^{\;c}K_{c}^{\;a}\right)  \right)  \right\}
\end{align}
Thus we reach the final result that the spectral action for the standard model
including all boundary terms is given by
\begin{align}
&  I=\frac{48\Lambda^{4}}{\pi^{2}}f_{4}%
%TCIMACRO{\dint \limits_{M}}%
%BeginExpansion
{\displaystyle\int\limits_{M}}
%EndExpansion
d^{4}x\sqrt{g}\nonumber\\
&  +\frac{8\Lambda^{2}}{\pi^{2}}f_{2}\left\{
%TCIMACRO{\dint \limits_{M}}%
%BeginExpansion
{\displaystyle\int\limits_{M}}
%EndExpansion
d^{4}x\sqrt{g}\left(  -\frac{1}{2}R-\frac{1}{4}\left(  a\left\vert
\varphi\right\vert ^{2}+\frac{1}{2}c\right)  \right)  \right. \nonumber\\
&  \qquad\qquad\left.  -%
%TCIMACRO{\dint \limits_{\partial M}}%
%BeginExpansion
{\displaystyle\int\limits_{\partial M}}
%EndExpansion
d^{3}x\sqrt{h}K\right\} \nonumber\\
&  +\frac{2\Lambda}{(4\pi)^{\frac{3}{2}}}f_{1}%
%TCIMACRO{\dint \limits_{\partial M}}%
%BeginExpansion
{\displaystyle\int\limits_{\partial M}}
%EndExpansion
d^{3}x\sqrt{h}\left(  3\left(  K^{2}-2K_{ab}K^{ab}\right)  \right) \nonumber\\
&  +\frac{f_{0}}{2\pi^{2}}\left\{
%TCIMACRO{\dint \limits_{M}}%
%BeginExpansion
{\displaystyle\int\limits_{M}}
%EndExpansion
d^{4}x\sqrt{g}\left(  -\frac{3}{5}C_{\mu\nu\rho\sigma}^{2}+\frac{11}%
{30}R^{\ast}R^{\ast}-(2/5)R_{;\mu}^{\,\,\mu}\right.  \right. \nonumber\\
&  \qquad\qquad+a\left\vert D_{\mu}\varphi\right\vert ^{2}+\frac{1}{6}R\left(
a\left\vert \varphi\right\vert ^{2}+\frac{1}{2}c\right) \nonumber\\
\qquad &  \qquad\qquad\left.  +g_{3}^{2}\left(  G_{\mu\nu}^{i}\right)
^{2}+g_{2}^{2}\left(  F_{\mu\nu}^{\alpha}\right)  ^{2}+\frac{5}{3}g_{1}%
^{2}\left(  B_{\mu\nu}\right)  ^{2}\right) \nonumber\\
&  \qquad\qquad\left.  +b\left\vert \varphi\right\vert ^{4}+2e\left\vert
\varphi\right\vert ^{2}+\frac{1}{2}d-\frac{1}{3}a\left(  \left\vert
\varphi\right\vert ^{2}\right)  _{;\mu}^{\,\,\mu}\right\} \nonumber\\
&  +\frac{f_{0}}{2\pi^{2}}\left\{
%TCIMACRO{\dint \limits_{\partial M}}%
%BeginExpansion
{\displaystyle\int\limits_{\partial M}}
%EndExpansion
d^{3}x\sqrt{h}\left(  \frac{1}{3}K\left(  a\left\vert \varphi\right\vert
^{2}+\frac{1}{2}c\right)  \right.  \right. \nonumber\\
&  \qquad\left.  +\frac{2}{15}\left(  5RK+4KR_{\;nan}^{a}+4K_{ab}R_{\;acb}%
^{c}+18R_{anbn}K^{ab}\right)  \right) \nonumber\\
&  \qquad+\frac{4}{315}\left(  17K^{3}+39KK_{ab}K^{ab}-116K_{a}^{\;b}%
K_{b}^{\;c}K_{c}^{\;a}\right)  ,
\end{align}
where
\begin{equation}
f_{n}=%
%TCIMACRO{\dint \limits_{0}^{\infty}}%
%BeginExpansion
{\displaystyle\int\limits_{0}^{\infty}}
%EndExpansion
v^{n-1}f(v)dv.
\end{equation}
There are two things to be noted about the form of the boundary terms. First,
the Higgs fields do contribute through the combination
\[
\frac{1}{3}K\left(  a\left\vert \varphi\right\vert ^{2}+\frac{1}{2}c\right)
.
\]
This is dictated by the presence of the term
\[
\frac{1}{6}R\left(  a\left\vert \varphi\right\vert ^{2}+\frac{1}{2}c\right)
\]
and therefore, they again appear together, with the same sign and relative
factor of 2. This is remarkable and means that the spectral action takes care
of its own consistency. The second thing, is the absence of the contributions
of the gauge fields to boundary terms. It is known that both in the
Hamiltonian formulation, or Lagrangian path integrals, a boundary term is
added to the make the definition of conjugate momenta possible and to enforce
the Gauss constraint on the divergence of the electric field. It was, however,
shown by Vassilevich \cite{Vass3} (section 3.4) and \cite{Vass2} that the
Yang-Mills action
\begin{equation}
\frac{1}{4}%
%TCIMACRO{\dint \limits_{M}}%
%BeginExpansion
{\displaystyle\int\limits_{M}}
%EndExpansion
d^{4}x\sqrt{g}\left(  F_{\mu\nu}^{\alpha}F^{\mu\nu\alpha}\right)
\end{equation}
where
\begin{equation}
F_{\mu\nu}^{\alpha}=\partial_{\mu}A_{\nu}^{\alpha}-\partial_{\nu}A_{\mu
}^{\alpha}+f_{\beta\gamma}^{\alpha}A_{\mu}^{\beta}A_{\nu}^{\gamma}%
\end{equation}
can be put into the form
\begin{align*}
&  \frac{1}{2}%
%TCIMACRO{\dint \limits_{M}}%
%BeginExpansion
{\displaystyle\int\limits_{M}}
%EndExpansion
d^{4}x\sqrt{g}A^{\rho\alpha}\left(  \left(  -g_{\rho\sigma}g^{\mu\nu}%
\nabla_{\mu}\nabla_{\nu}+\nabla_{\rho}\nabla_{\sigma}+R_{\rho\sigma}\right)
g_{\alpha\beta}+2F_{\rho\sigma}^{\gamma}\left(  B\right)  f_{\alpha\beta
}^{\gamma}\right)  A^{\sigma\beta}\\
&  +\frac{1}{2}%
%TCIMACRO{\dint \limits_{\partial M}}%
%BeginExpansion
{\displaystyle\int\limits_{\partial M}}
%EndExpansion
d^{3}x\sqrt{h}A^{\nu\alpha}\left(  \nabla_{n}A_{\nu\alpha}-\nabla_{\nu
}A_{n\alpha}\right)
\end{align*}
where the field $A_{\mu}^{\alpha}$ is expanded around a background $B_{\mu
}^{\alpha}$. By imposing the gauge condition
\begin{equation}
\nabla^{\mu}A_{\mu}^{\alpha}=0
\end{equation}
and one of the two boundary conditions
\begin{equation}
A_{n}|_{\partial M}=0,\quad\left(  \nabla_{n}\delta_{ab}-K_{ab}\right)
A_{b}|_{\partial M}=0
\end{equation}
or
\begin{equation}
\left(  \nabla_{n}-K\right)  A_{n}|_{\partial M}=0,\quad A_{a}|_{\partial M}=0
\end{equation}
the boundary term will vanish in both cases. We have noted that we have taken
the first condition to avoid any appearance of gauge fields in the boundary.
In other words the spectral action needs only part of the first two
conditions. It also seems to imply that the second part of the boundary
conditions arise as integrability condition derived from the boundary
conditions of the Dirac operator.

From all these considerations we deduce that the simple requirement of having
boundary conditions for the Dirac operator which are consistent with the
self-adjointness of this operator, is enough to guarantee that the spectral
action has all the correct boundary terms, including correct signs and coefficients.

\section{Spectral action in presence of dilaton}

We now deal with the question of what is the form of the action if a scaling
is introduced through the operator
\begin{equation}
e^{-\phi}D^{2}e^{-\phi}=-\left(  G^{\mu\nu}\partial_{\mu}\partial_{\nu
}+\mathcal{A}^{\mu}\partial_{\mu}+\mathcal{B}\right)
\end{equation}
where
\begin{align*}
G^{\mu\nu}  &  =e^{-2\phi}g^{\mu\nu},\\
\mathcal{A}^{\mu}  &  =e^{-2\phi}A^{\mu}-2G^{\mu\nu}\partial_{\nu}\phi,\\
\mathcal{B}  &  =e^{-2\phi}B+G^{\mu\nu}\left(  \partial_{\mu}\phi\partial
_{\nu}\phi-\partial_{\mu}\partial_{\nu}\phi\right)  -e^{-2\phi}A^{\mu}%
\partial_{\mu}\phi.
\end{align*}
We have shown in \cite{scale} that for this operator we have the identity
\begin{equation}
\mathcal{E}+\frac{1}{6}R\left(  G\right)  =e^{-2\phi}\left(  E+\frac{1}%
{6}R\left(  g\right)  \right)
\end{equation}
where
\begin{align*}
\mathcal{E}  &  =\mathcal{B}-G^{\mu\nu}\left(  \partial_{\mu}\overline
{\omega^{\prime}}_{\nu}+\overline{\omega^{\prime}}_{\mu}\overline
{\,\omega^{\prime}}_{\nu}-\Gamma_{\mu\nu}^{\rho}\left(  G\right)
\overline{\omega^{\prime}}_{\rho}\right)  ,\\
\overline{\omega^{\prime}}_{\mu}  &  =\frac{1}{2}G_{\mu\nu}\left(
\mathcal{A}^{\nu}+\Gamma^{\nu}\left(  G\right)  \right)  ,\\
\mathbf{\Omega}_{\mu\nu}  &  =\partial_{\mu}\overline{\omega^{\prime}}_{\nu
}-\partial_{\nu}\overline{\omega^{\prime}}_{\mu}+\left[  \overline
{\omega^{\prime}}_{\mu},\overline{\omega^{\prime}}_{\nu}\right]  .
\end{align*}
It is then convenient to use these relations as well as
\begin{equation}
R\left(  g\right)  =e^{2\phi}\left(  R\left(  G\right)  -6\,G^{\mu\nu}\left(
-\nabla_{\mu}^{G}\nabla_{\nu}^{G}\phi+\partial_{\mu}\phi\partial_{\nu}%
\phi\right)  \right)  ,
\end{equation}
to work out the spectral action for the scaled operator on manifolds with
boundary. A good starting point is the equality
\begin{equation}
\left\langle \Psi\left\vert D\right\vert \Psi\right\rangle =\,\left\langle
\Psi^{\prime}\left\vert D^{\prime}\right\vert \Psi^{\prime}\right\rangle
^{\prime}%
\end{equation}
where
\begin{equation}
\left\vert \Psi\right\rangle =e^{\frac{3}{2}\phi}\left\vert \Psi^{^{\prime}%
}\right\rangle
\end{equation}
then the boundary conditions are taken to be
\begin{equation}
\Pi_{-}\Psi^{^{\prime}}|_{\partial M}=0,
\end{equation}
which implies the boundary condition for $D^{2}$
\begin{equation}
\Pi_{-}D^{^{\prime}}\Psi^{^{\prime}}|_{\partial M}=0
\end{equation}
so that the function $S$ is evaluated using the rescaled metric $G_{\mu\nu}$.
To cut the story short, there are only few places where we expect the dilaton
to contribute. The terms in the bulk have already been evaluated, except for
the total divergence $\left(  5E+R\right)  _{;\mu}^{\,\,\mu}$ which does
receive a dilaton contribution equal to
\begin{align}
&  \frac{5}{2}%
%TCIMACRO{\dint \limits_{M}}%
%BeginExpansion
{\displaystyle\int\limits_{M}}
%EndExpansion
d^{4}x\sqrt{G}\left[  G^{\kappa\lambda}\left(  \partial_{\kappa}\phi
\partial_{\lambda}\phi-\nabla_{\kappa}^{G}\nabla_{\lambda}^{G}\phi\right)
\right]  _{;\mu}^{\;\mu}\\
&  =\frac{5}{2}%
%TCIMACRO{\dint \limits_{M}}%
%BeginExpansion
{\displaystyle\int\limits_{M}}
%EndExpansion
d^{3}y\sqrt{H}\partial_{n}\left[  H^{ab}\left(  \partial_{a}\phi\partial
_{b}\phi-\nabla_{a}^{H}\nabla_{b}^{H}\phi\right)  +\left(  \partial_{n}%
\phi\partial_{n}\phi-\nabla_{n}\nabla_{n}\phi\right)  \right]
\end{align}
The remaining boundary terms could be simplified by observing that first
modification occurs for $a_{4}\left(  e^{-\phi}D^{2}e^{-\phi},\chi\right)  $
where we have the combination
\begin{equation}
720\text{\textrm{Tr}}\left(  \mathcal{E}+\frac{1}{6}R\left(  G\right)
\right)  \left(  S+\frac{1}{6}K\right)
\end{equation}
which, in the case of the standard model, is equal to%
\begin{align*}
&  \frac{1}{16\pi^{2}}\frac{1}{360}f_{0}%
%TCIMACRO{\dint \limits_{\partial M}}%
%BeginExpansion
{\displaystyle\int\limits_{\partial M}}
%EndExpansion
d^{3}y\sqrt{H}720\left(  -\frac{1}{12}K\right)  \left(  384\right)  \left(
-\frac{1}{12}\right) \\
&  \qquad\qquad\left(  R\left(  G\right)  +6\,H^{ab}\left(  \nabla_{a}%
^{H}\nabla_{b}^{H}\phi-\partial_{a}\phi\partial_{b}\phi\right)  +6\left(
\nabla_{n}\nabla_{n}\phi-\partial_{n}\phi\partial_{n}\phi\right)  \right)
\end{align*}
which implies that the last term for $a_{4}$ gets modified by replacing
$\frac{1}{8}\left(  5RK+\cdots\right)  $ by
\begin{equation}
\frac{1}{3\pi^{2}}f_{0}%
%TCIMACRO{\dint \limits_{\partial M}}%
%BeginExpansion
{\displaystyle\int\limits_{\partial M}}
%EndExpansion
d^{3}y\sqrt{H}K\left(  R\left(  G\right)  +6H^{ab}\left(  \nabla_{a}^{H}%
\nabla_{b}^{H}\phi-\partial_{a}\phi\partial_{b}\phi\right)  +6\left(
\nabla_{n}\nabla_{n}\phi-\partial_{n}\phi\partial_{n}\phi\right)  \right)
\end{equation}
and the boundary term, not being conformally invariant, gets a contribution
dependent on the dilaton. Therefore the full action takes exactly the same
form as before, but as function of the metric $G_{\mu\nu}$ and the induced
metric $H_{ab}=e^{2\phi}h_{ab}$ and the Higgs field $\varphi^{\prime}%
=e^{-\phi}\varphi$ and the fermions $\Psi^{\prime}=e^{-\frac{3}{2}\phi}\Psi,$
plus the extra terms
\begin{align}
&  \frac{12}{\pi^{2}}f_{2}\int_{M}d^{4}x\sqrt{G}G^{\mu\nu}\partial_{\mu}%
\phi\partial_{\nu}\phi\\
&  +\frac{2}{\pi^{2}}f_{0}%
%TCIMACRO{\dint \limits_{\partial M}}%
%BeginExpansion
{\displaystyle\int\limits_{\partial M}}
%EndExpansion
d^{3}x\sqrt{H}\left(  K+\partial_{n}\right)  \left[  H^{ab}\left(  \nabla
_{a}^{H}\nabla_{b}^{H}\phi-\nabla_{a}\phi\nabla_{b}\phi\right)  +\left(
\nabla_{n}\nabla_{n}\phi-\partial_{n}\phi\partial_{n}\phi\right)  \right]
\end{align}
Practical applications of these results will be dealt with in the future.

\section{Appendix 1: the case of the disk}

We take the case of the Dirac operator in the unit disk, in order to check the
conventions for the extrinsic curvature. We take the Dirac operator in the
form:
\[
D=\left(
\begin{array}
[c]{cc}%
0 & \partial_{x}+i\partial_{y}\\
-\partial_{x}+i\partial_{y} & 0
\end{array}
\right)
\]
and we write it in polar coordinates $(r,\theta)$ using
\[
\partial_{x}=\cos\theta\,\partial_{r}-\sin\theta\frac{1}{r}\partial_{\theta
}\,,\ \ \partial_{y}=\sin\theta\,\partial_{r}+\cos\theta\frac{1}{r}%
\partial_{\theta}%
\]
so that
\[
D=i(\gamma_{1}(\theta)\frac{1}{r}\,\partial_{\theta}+\gamma_{2}(\theta
)\,\partial_{r})
\]
where
\[
\gamma_{1}(\theta)=\left(
\begin{array}
[c]{cc}%
0 & e^{i\theta}\\
e^{-i\theta} & 0
\end{array}
\right)  \,,\ \ \gamma_{2}(\theta)=\left(
\begin{array}
[c]{cc}%
0 & -ie^{i\theta}\\
ie^{-i\theta} & 0
\end{array}
\right)  \,.
\]
The $\gamma_{j}(\theta)$ are self-adjoint of square $1$ and fulfill the
Clifford relations
\[
\gamma_{1}(\theta)\gamma_{2}(\theta)=-\gamma_{2}(\theta)\gamma_{1}%
(\theta)=i\,\gamma
\]
where $\gamma=\left(
\begin{array}
[c]{cc}%
1 & 0\\
0 & -1
\end{array}
\right)  $ gives the grading. Note also that
\[
\partial_{\theta}(\gamma_{1}(\theta))=-\gamma_{2}(\theta)
\]

\begin{lemma}
The boundary condition for $D$ is given by
\[
\gamma_{1}(\theta)\,\xi=\xi
\]

\end{lemma}

\proof By definition the boundary condition is given as $\Pi_{-}\xi=0$ on the
boundary, where (\cite{Vass3} p. 297)
\[
\Pi_{-}=\frac{1}{2}(1-i\gamma_{n}\gamma)
\]
with $\gamma_{n}$ the Clifford multiplication by the normal, and $\gamma$ the
grading as above.

We have $D=i(\gamma_{x}\partial_{x}+\gamma_{y}\partial_{y})$ where
\[
\gamma_{x}=\left(
\begin{array}
[c]{cc}%
0 & -i\\
i & 0
\end{array}
\right)  \,,\ \ \gamma_{y}=\left(
\begin{array}
[c]{cc}%
0 & 1\\
1 & 0
\end{array}
\right)
\]
and thus the inward normal corresponds to
\[
\gamma_{n}=-\cos\theta\gamma_{x}-\sin\theta\gamma_{y}=-\gamma_{2}(\theta)\,.
\]
One has $i\gamma_{n}\gamma=-i\gamma_{2}(\theta)\gamma=\gamma_{1}(\theta)$ and
thus $\Pi_{-}=\frac{1}{2}(1-\gamma_{1}(\theta))$. \endproof

\begin{lemma}
The additional boundary condition for $D^{2}$ is given by
\[
(\partial_{n}-\frac12)\Pi_{+}\xi=0\,,\ \ \Pi_{+}=\frac12(1+\gamma_{1}%
(\theta))
\]
where $\partial_{n}=-\partial_{r}$ is differentiation relative to the inward normal.
\end{lemma}

\proof The additional boundary condition is
\[
\Pi_{-}\,D\,\xi=0\,, \ \ \Pi_{-}=\frac12(1-\gamma_{1}(\theta))\,.
\]
Up to an overall factor this gives $N\xi=0$ with
\[
N=(1-\gamma_{1}(\theta))(\gamma_{1}(\theta)\frac1r\,\partial_{\theta}+
\gamma_{2}(\theta)\,\partial_{r})\,.
\]
One has
\[
(1-\gamma_{1}(\theta)) \gamma_{2}(\theta)\,\partial_{r}= \gamma_{2}%
(\theta)\,\partial_{r}\,(1+\gamma_{1}(\theta))=2\gamma_{2}(\theta
)\,\partial_{r}\,\Pi_{+}\, .
\]
Next, using the first boundary condition, one gets $\partial_{\theta}%
(1-\gamma_{1}(\theta))\xi=0$ on the boundary circle. One has moreover
\[
\partial_{\theta}\xi=\partial_{\theta}(\gamma_{1}(\theta)\xi)=(\partial
_{\theta}(\gamma_{1}(\theta))\xi+ \gamma_{1}(\theta)\partial_{\theta}%
\xi=-\gamma_{2}(\theta)\xi+ \gamma_{1}(\theta)\partial_{\theta}\xi
\]
Thus on the boundary one has
\[
(1-\gamma_{1}(\theta))\partial_{\theta}\xi=-\gamma_{2}(\theta)\xi\,.
\]
This yields, on the boundary,
\[
N=\gamma_{2}(\theta)\frac1r+2\gamma_{2}(\theta)\,\partial_{r}\,\Pi_{+}
\]
and hence
\[
-\frac12\,\gamma_{2}(\theta)\,N=(\partial_{n}-\frac12)\Pi_{+}
\]
since $\partial_{n}=-\partial_{r}$. \endproof

\medskip

\section{Appendix 2: Sign of boundary term in Einstein action}

We use the notations of \cite{Poisson}. We check that in Euclidean signature
the correct combination which gives the Einstein equation is
\[
-\int_{M} R \sqrt g d^{4}x - 2 \int_{\partial M} K \sqrt h d^{3}y
\]
where $R$ is positive for the sphere, and $K$ is positive for the ball. This
fits with Hawking \cite{Hawking} from which one can also check that the
Euclidean action is as above for the overall sign.

\subsection{Sign of $R$}

The Ricci scalar is defined by
\begin{equation}
R=g^{\mu\nu}\,R_{\mu\nu},\ R_{\mu\nu}=R_{\ \mu\rho\nu}^{\rho} \label{defofr}%
\end{equation}
where in a geodesic coordinate system
\begin{equation}
R_{\mu\nu\rho\sigma}=\frac{1}{2}(g_{\mu\sigma,\nu\rho}-g_{\mu\rho,\nu\sigma
}-g_{\nu\sigma,\mu\rho}+g_{\nu\rho,\mu\sigma}) \label{defofr1}%
\end{equation}
Thus for the sphere with $g_{\mu\nu}=(1+\frac{\Omega}{4}\rho^{2})^{-2}%
\delta_{\mu\nu}$ the value of $R$ is $\frac{n(n-1)}{2}\,\Omega$ which is
positive since $\Omega>0$.

\subsection{Stokes formula and outer normal}

We start with a vector field $X=X^{\mu}\partial_{\mu}$ on a manifold with
volume form $\omega$. The divergence of $X$ is given by
\[
\mathrm{div}X=d\,i_{X}\omega
\]
which is the Lie derivative $\partial_{X}\omega$ of the volume form since
$\partial_{X}=di_{X}+i_{X}d$ on forms. The Stokes formula gives
\[
\int_{M}\,\mathrm{div}X=\int_{\partial M}\,i_{X}\omega
\]
where both $M$ and $\partial M$ are \emph{oriented} so that
\[
\int_{M}\,d\alpha=\int_{\partial M}\alpha
\]
Using a Riemannian metric $g$ on $M$ and the induced metric $h$ on $\partial
M$ we get a formula of the form
\begin{equation}
\int_{M}\,X_{\ ;\,\mu}^{\mu}\sqrt{g}d^{n}x=-\int_{\partial M}\,X^{\mu}%
\,n_{\mu}\sqrt{h}d^{n-1}y \label{stokes}%
\end{equation}
and we need to determine the sign of the normal $n_{\mu}=g_{\mu\nu}n^{\nu}$.
Note that in this formula the choice of orientation of $M$ has disappeared. To
get the sign of $n^{\nu}$ one can take the one dimensional case where
$M=[a,b]$ with $a<b$. Thus the coordinate $x$ increases from $a$ to $b$. One
lets $X=f(x)\partial_{x}$. The left hand side of \eqref{stokes} gives
\[
\int_{a}^{b}\,\partial_{x}f(x)dx=f(b)-f(a)
\]
which shows that $n^{\nu}$ is the \emph{inward} normal. More generally if we
let $k(x)$ be a convex function such as $k(x)=\sum(x^{\mu})^{2}$ in
$\mathbb{R}^{n}$ and take $M=\{x|\,k(x)\leq1\}$, we can take for $X$ the
gradient of $k$. Then the left hand side of \eqref{stokes} is positive and
thus the normal is again the \emph{inward} normal.

\subsection{Extrinsic curvature}

We now recall the definition of the extrinsic curvature \cite{Poisson}: The
\emph{extrinsic curvature} $K_{ab}$ is defined by
\[
K_{ab}=-n_{\mu;\,\nu}\,e_{a}^{\mu}e_{b}^{\nu}%
\]
where $n^{\mu}$ is the \emph{inward } normal.

Let us compute it explicitly in the case of the disk of radius $R$ in the
plane with coordinates $x^{\mu}$ and flat metric $g_{\mu\nu}=\delta_{\mu\nu}$.
We take for $y$ the angular parameter $y=\theta$ so that
\[
x^{1}(y)=R\cos\theta\,,\ x^{2}(y)=R\sin\theta
\]
There is only one index $a=1$ and one has
\[
e_{1}^{1}=\partial_{\theta}R\cos\theta=-R\sin\theta\,,\ \ e_{1}^{2}%
=\partial_{\theta}R\sin\theta=R\cos\theta\,.
\]
The coordinates of the inward normal are
\[
n^{\mu}=-x^{\mu}/\sqrt{(x^{1})^{2}+(x^{2})^{2}}%
\]
One finds by direct computation that
\[
-n_{\mu;\,\nu}\,e_{1}^{\mu}e_{1}^{\nu}=\sqrt{(x^{1})^{2}+(x^{2})^{2}}=R
\]
One has $h_{11}=R^{2}$ and thus $h^{11}=R^{-2}$ which gives in this case
\[
K=h^{ab}K_{ab}=\frac{1}{R}\,.
\]
One defines $h^{\mu\nu}$ by
\begin{equation}
h^{\mu\nu}=h^{ab}e_{a}^{\mu}e_{b}^{\nu} \label{halphabeta}%
\end{equation}
then
\[
K=h^{ab}K_{ab}=-h^{ab}n_{\mu;\,\nu}\,e_{a}^{\mu}e_{b}^{\nu}=-h^{\mu\nu}%
n_{\mu;\,\nu}.
\]

\subsection{Variation of the Einstein action}

This is well known, but to fix the notation, we give the main steps. The
intermediate steps are given in \cite{Poisson}.

Variation of the Einstein action is%

\begin{align*}
\delta I_{E}  &  =\int_{M}\,\delta(g^{\mu\nu}R_{\mu\nu}\sqrt{g})d^{4}x\\
&  =\int_{M}\,(R_{\mu\nu}-\frac{1}{2}Rg_{\mu\nu})\delta g^{\mu\nu}\sqrt
{g}d^{4}x+\int_{M}\,g^{\mu\nu}\delta R_{\mu\nu}\sqrt{g}d^{4}x
\end{align*}
We then use
\begin{equation}
g^{\mu\nu}\delta R_{\mu\nu}=X_{\;;\,\mu}^{\mu}%
\end{equation}
where
\[
X^{\mu}=g^{\nu\rho}\delta\Gamma_{\nu\rho}^{\mu}-g^{\nu\mu}\delta\Gamma
_{\alpha\rho}^{\rho}%
\]
Using Stokes theorem
\begin{align}
\int_{M}X_{\;;\,\mu}^{\mu}\sqrt{g}d^{4}x  &  =-\int_{\partial M}n^{\mu}X_{\mu
}\sqrt{h}d^{3}y\\
&  =-\int_{\partial M}h^{\mu\nu}(\delta g_{\rho\nu,\,\mu}-\delta g_{\mu
\nu,\,\rho})n^{\rho}\sqrt{h}d^{3}y\\
&  =\int_{\partial M}h^{\mu\nu}\delta g_{\mu\nu,\,\rho}n^{\rho}\,\sqrt{h}%
d^{3}y
\end{align}
where in the last step we used that the variation of $g_{\mu\nu}$ and the
tangential derivative of $\delta g_{\mu\nu}$is zero on $\partial M$ so that
$\delta g_{\mu\nu,\,\alpha}e_{a}^{\alpha}=0$ and $h^{\alpha\beta}\delta
g_{\mu\nu,\,\alpha}=0.$

For the variation of the boundary term we have
\begin{align}
\delta\int_{\partial M}2K\sqrt{h}d^{3}y  &  =\int_{\partial M}2h^{\mu\nu
}\delta\Gamma_{\,\mu\nu}^{\rho}n_{\rho}\sqrt{h}d^{3}y\\
&  =-\int_{\partial M}h^{\mu\nu}\delta g_{\mu\nu,\,\rho}n^{\rho}\sqrt{h}d^{3}y
\end{align}
Thus
\begin{equation}
\delta\left(
%TCIMACRO{\dint \limits_{M}}%
%BeginExpansion
{\displaystyle\int\limits_{M}}
%EndExpansion
d^{4}x\sqrt{g}R+2%
%TCIMACRO{\dint \limits_{\partial M}}%
%BeginExpansion
{\displaystyle\int\limits_{\partial M}}
%EndExpansion
d^{3}x\sqrt{h}K\right)  =\int_{M}\,(R_{\mu\nu}-\frac{1}{2}Rg_{\mu\nu})\delta
g^{\mu\nu}\sqrt{g}d^{4}x
\end{equation}

\section*{Acknowledgment}

The research of A. H. C. is supported in part by the National Science
Foundation under Grant No. Phys-0854779.


\begin{thebibliography}{99}                                                                                               %


\bibitem {Barth}N. Barth, ``The fourth order gravitational action for
manifolds with boundaries", \textit{Class. Quant. Grav. }\textbf{2} (1985) 497.

\bibitem {BG1}T. Branson and P. Gilkey, ``Residues for the eta function for an
operator of Dirac type with local boundary conditions", \textit{Diff. Geom.
Appl. }\textbf{2 }(1992) 249.

\bibitem {BG2}T. Branson and P. Gilkey, ``Residues of the eta function for an
operator of Dirac type", \textit{Journal of Functional Analysis }\textbf{108
}(1992) 47.

\bibitem {BGV}T. Branson, P. Gilkey and D. Vassilevich, ``Vacuum expectation
value asmyptotics for second order differential operators on manifolds with
boundary", \textit{J. Math. Phys. }\textbf{39 }(1998) 1040.

\bibitem {ACAC}Ali H. Chamseddine and Alain Connes, \textquotedblleft
Universal Formula for Noncommutative Geometry Actions: Unification of Gravity
and the Standard Model", \textit{Phys. Rev. Lett. }\textbf{77 }4868 (1996);
"The Spectral Action Principle" \textit{Comm. Math. Phys. }\textbf{186 }731 (1997).

\bibitem {ACM}A.~H.~Chamseddine, A.~Connes and M.~Marcolli, \emph{Gravity and
the standard model with neutrino mixing}
Adv.\ Theor.\ Math.\ Phys.\ \textbf{11}, 991 (2007) [arXiv:hep-th/0610241].

\bibitem {scale}Ali H. Chamseddine and Alain Connes, \textquotedblleft Scale
Invariance in the Spectral Action" \textit{J. Math. Phys. }\textbf{47
}063504\textbf{ }(2006).

\bibitem {Beggar}A.~H.~Chamseddine and A.~Connes, \emph{Conceptual Explanation
for the Algebra in the Noncommutative Approach to the Standard Model}
Phys.\ Rev.\ Lett.\ \textbf{99}, 191601 (2007) [arXiv:0706.3690 [hep-th]]

\bibitem {Why}A.~H.~Chamseddine and A.~Connes, \emph{Why the Standard Model}
J.\ Geom.\ Phys.\ \textbf{58}, 38 (2008) [arXiv:0706.3688 [hep-th]].

\bibitem {quantumactions}Ali H. Chamseddine and Alain Connes,
\textquotedblleft Quantum Gravity Boundary Terms from Spectral Action",
\textit{Phys. Rev. Lett. }\textbf{99 }071302 (2007), arXiv:0705.1786.

\bibitem {Gilkeyb1}P. Gilkey ``Invariance Theory, the heat equation and the
Atiyah-Singer Index theorem", CRC press, second edition.

\bibitem {Gilkeyb2}P. Gilkey ``Asymptotic Formulae in Spectral Geometry", CRC
press, 2004.

\bibitem {GK}P. Gilkey and K. Kristen, ``Stability theorems for chiral bag
boundary conditions", \textit{Lett. Math. Phys. }\textbf{73 }(2005) 147.

\bibitem {Hawking}Editors G. Gibbons and S. Hawking, ``Euclidean Quantum
Gravity", World Scientific 1993.

\bibitem {Kuchar}K. Kuchar, ``Geometry of hyperspace I and II, \textit{J.
Math. Phys. }\textbf{17 }(1976) 777 and 792.

\bibitem {Luck}H. Luckock ``Mixed boundary conditions in quantum field
theory", \textit{J. Math. Phys. }\textbf{32 }(1991) 1755.

\bibitem {MTW}C. Misner K. Thorne and J. Wheeler "Gravitation" section 21.4-21.8

\bibitem {Poisson}E. Poisson, ``An advanced course in General Relativity", 2002.

\bibitem {Vass2}D. Vassilevich, ``The Faddeev-Popov trick in the presence of
boundaries", \textit{Phys. Lett. }\textbf{B421 }(1998) 93.

\bibitem {Vass3}D. Vassilevich, \textquotedblleft Heat kernel expansion:
user's manual", \textit{Physics Reports } \textbf{388 }(2003) 279-360.
\end{thebibliography}
\end{document}